\documentclass[12pt]{article} 
\usepackage{graphicx,floatflt,amssymb,rotate}
\usepackage{mathrsfs}
\usepackage{amssymb}
\usepackage{amsmath}
\usepackage{color}
\usepackage{graphicx}
\usepackage{subfigure}
\usepackage{multirow}
\usepackage{float}
\usepackage{pstricks}
\usepackage[toc,page]{appendix}
\usepackage[bottom=.1in]{geometry}

\begin{document}
\vskip 30pt  
 
\begin{center}  
{\Large \bf Non-minimal UED confronts \boldmath{$B_{s}\rightarrow\mu^{+}\mu^{-}$} } \\
\vspace*{1cm}  
\renewcommand{\thefootnote}{\fnsymbol{footnote}}  
{ {\sf Anindya Datta \footnote{email: adphys@caluniv.ac.in}},  
{\sf Avirup Shaw \footnote{email: avirup.cu@gmail.com}} 
} \\  
\vspace{10pt}  
{ {\em Department of Physics, University of Calcutta,  
92 Acharya Prafulla Chandra Road, \\ Kolkata 700009, India}}
\normalsize  
\end{center} 

\begin{abstract}
Addition of boundary localised kinetic and Yukawa terms to the action of a 5-dimensional Standard Model would non-trivially modify the Kaluza-Klein spectra 
and some of the interactions among the Kaluza-Klein excitations 
compared to the minimal version of this model, in which, these boundary terms are not present.  In the minimal version of this framework known as Universal Extra Dimensional model, special assumptions 
are made about these unknown, beyond the cut-off contributions to restrict  the number of unknown parameters of the theory to a minimal.  
We estimate the contribution of Kaluza-Klein modes to the branching 
ratios of $B_{s(d)}\rightarrow\mu^{+}\mu^{-}$  in the framework of non-minimal Universal Extra Dimensional, at one loop level. The results have been compared to the experimental data to constrain the parameters of this model.
 From the measured decay branching ratio of $B_s \rightarrow \mu^+ \mu^-$ (depending on the values of boundary localised parameters) lower limit on $R^{-1}$ can be as high as 800 GeV.  We have briefly reviewed the bounds on nmUED parameter space coming from electroweak precision observables. The present analysis ($B_s \rightarrow \mu^+ \mu^-$) has ruled out new regions of parameter space 
 in comparison to the analysis of electroweak data. We have revisited the bound on $R^{-1}$ in Universal Extra Dimensional model, which came out to be 454 GeV.  This limit on $R^{-1}$ in Universal Extra Dimensional framework is not as competitive 
 as the limits derived from the consideration of relic density or Standard Model Higgs boson production and decay to $W^+ W^-$.
 Unfortunately, $B_{d}\rightarrow\mu^{+}\mu^{-}$ decay branching ratio would not set any significant limit
on $R^{-1}$ in a minimal or non-minimal Universal Extra Dimensional model.
 
\end{abstract}
\noindent PACS No: {\tt 11.10 Kk, 12.60.-i, 14.70.Hp, 14.80.Rt}\renewcommand{\thesection}{\Roman{section}}  
\setcounter{footnote}{0}  
\renewcommand{\thefootnote}{\arabic{footnote}}

\section{Introduction}

After the discovery of the Higgs boson at the LHC experiment, the new challenge to particle physics is to provide a framework in which there exists a natural Dark Matter (DM) candidate, as the Standard Model (SM) itself does not 
have a sufficiently massive weakly interacting particle to be a good candidate for DM. Thus one is compelled to look beyond the SM and in this endeavour, extra dimensional scenarios offer such a paradigm. Some of the variants of extra dimensional theories offer the solution to the  DM \cite{ued_dm, relic} puzzle along with many others like gauge coupling unifications \cite{ued_uni} and  fermion mass hierarchy \cite{hamed} to name a few. A particular extension  of the SM needs special attention in this regard.
This is known as Universal Extra Dimensional (UED) Model where all the SM fields can propagate in $4 +1$ dimensional space-time. The extra dimension (say, $y$) is compactified on a circle ($S^{1}$)  of radius $R$ \cite{acd}. 5-dimensional (5-D) action, which has the same field content as  the SM, would respect the same $SU(3)_c \times SU(2)_L \times U(1)_Y$ gauge symmetry.
The 4-dimensional (4-D) effective theory is characterised by Kaluza-Klein (KK) towers corresponding to each SM field.
The mass of a $n^{th}$ KK-mode excitation is given by $m_n^2 = m^2 + \frac{n^2}{R^2}$; $n$ being the KK-number which is
nothing but the discretized momentum in $y$-direction  and $m$ is the 0-mode mass. SM particles have been identified with the $n=0$ mode fields in the effective theory. However there is one caveat, unlike the SM, the 0-mode fermions in the effective theory are not chiral, but are vector like in nature.   
To get rid of the unwanted fermion 0-modes, one needs to impose an extra $Z_{2}$ symmetry: $y \rightarrow -y$ on the action. 
Fields which do not have 0-modes are chosen to be odd under this $Z_{2}$ symmetry. Rest of the fields, having 0-modes,  are even under this discrete 
transformation. The space of $y$ is restricted from $0$ to $\pi R$ and is  called an $S^{1}/Z_{2}$ orbifold. 
Two boundary points of the orbifold are also called the fixed points as they transform onto themselves under this $Z_2$ transformation.

Radiative correction plays \cite{rad_cor_georgi, rad_cor_cheng} an important role to cure a highly degenerate KK-mass spectrum in this model. These corrections fall into
two categories: namely the finite bulk corrections and more importantly the boundary localised corrections 
having logarithmic dependence on the cut-off scale $\Lambda$. A very special assumption is being taken in minimal UED (mUED) which ensures vanishing radiative corrections at the scale
 $\Lambda$. In a more general scenario like non-minimal UED (nmUED) \cite{Dvali}-\cite{ddrs1}, this special assumption is being avoided. Instead, 
 one assumes the boundary localised corrections as free parameters of the model. 
In this article, phenomenology of a particular non-minimal scenario in which  kinetic and Yukawa terms involving fields  are added to the 5-D action,
at boundary points, will be investigated. 
Coefficients of boundary localised kinetic terms (BLKT) and boundary localised Yukawa terms (BLYT) along with the radius of compactification, 
$R$ can be chosen as free parameters and experimental data can be used to constrain them. 

Various phenomenological studies in the framework of nmUED have been made to constrain non-minimality parameters from the perspective of 
electroweak observables \cite{flacke}, S, T and U parameters \cite{delAguila_STU, flacke_STU}, relic density 
\cite{tommy, ddrs2}, measurement of decay width of $Z$-boson to a pair of $b$-quarks \cite{zbb}, SM Higgs boson production and decay \cite{tirtha} and from the LHC experiments \cite{asesh_lhc1, lhc}.

Historically any search of {\em new physics} beyond the SM has been guided by the effects like precision electroweak variables like $\rho$(T)-parameter,
$R_b$ ($Z$-boson decay width to a pair of $b$ quarks normalised to total hadronic decay width), $A_{FB}^b$ (forward-backward asymmetry of $b$ quarks at $Z$-pole) etc.
Incidentally, all of these electroweak observables are extremely sensitive to the 
quantum corrections. Furthermore, large top quark mass plays a crucial role to amplify these quantum effects. 
In the same spirit, we would like to investigate how a precisely measured quantity namely, branching ratio of  $B_{s(d)}$ meson decay to $\mu^+ \mu^-$ could 
help us in this endeavour.  
Recent experimental measurement of these branching ratios by CMS \cite{cms} and LHCb \cite{lhcb} collaborations have created some excitements among the high
energy physics community, as the $B_s \rightarrow \mu^+ \mu^-$ decay has been thought to be one of the harbingers of new physics.
Unfortunately, the measured value of the branching ratio is more or less consistent with the SM prediction \cite{sm}. There exists only a one (two)-standard deviation difference between the experimentally measured value and the SM estimation of the $B_s$ ($B_d$) branching ratio
to $\mu^+ \mu^-$ pair. However, this close agreement of experiment with the SM, can be exploited to constrain the parameters of any BSM scenario. In this article we will be performing such an exercise. We will evaluate the $Br (B_{s(d)}\rightarrow \mu^+ \mu^-)$ in nmUED framework and compare our results with that from the experiment.

$Br (B_{s(d)}\rightarrow \mu^+ \mu^-)$  in the framework of UED has been previously estimated in ref.\;\cite{buras}.  
However, presence of boundary localised terms (BLT) in the action would non-trivially change  masses of KK-excitations  and  some of the couplings  
involving KK-excitations in nmUED framework.  Thus it would not be a trivial rescaling of an earlier calculation \cite{buras}
of $B_{s(d)}\rightarrow \mu^+ \mu^-$. To the best of our knowledge, this is the first estimation of $B_{s(d)}$ branching ratios in the framework of nmUED.

In the following section, we will evaluate relevant interactions and vertices in the framework of KK-parity-conserving nmUED with a brief
introduction of 
the model. In section 3 and 4 we will present some calculational details and numerical results respectively. The electroweak precision variables have been always very effective in constraining any form of new physics. We will briefly review the 
effect of precision observables (namely the $S$, $T$ and $U$) on the nmUED parameter space at the end.
We conclude in section 5.

\section{A lightning review of KK-parity-conserving nmUED scenario}

In this section, we discuss the KK-parity conserving nmUED model in brief, with focus on the interactions necessary for our calculations. 
We will need gauge and Yukawa interactions of the fermion KK-modes. Furthermore, we will briefly touch the spectrum of Goldstone bosons and physical
scalars as they play an important role in our analysis as we proceed to present a one loop calculation done in Feynman gauge, which is 
characterised by the presence of unphysical degrees of freedom in the particle spectra.

Let us start with the 5-D fermionic action with BLKT \cite{schwinn, ddrs1}:
\begin{eqnarray} 
S_{fermion} = \int d^5x \left[ \bar{\Psi}_L i \Gamma^M D_M \Psi_L 
+ r_f\{\delta(y)+\delta(y - \pi R)\} \bar{\Psi}_L i \gamma^\mu D_\mu P_L\Psi_L  
\right. \nonumber \\
\left. + \bar{\Psi}_R i \Gamma^M D_M \Psi_R
+ r_f\{\delta(y)+\delta(y - \pi R)\}\bar{\Psi}_R i \gamma^\mu D_\mu P_R\Psi_R
\right]. 
\label{factn}
\end{eqnarray}

In 5-dimensions four component Dirac spinors $\Psi_L(x,y)$ and $\Psi_R(x,y)$ can be written in terms of two component spinors \cite{schwinn, ddrs1}:
\begin{equation} 
\Psi_L(x,y) = \begin{pmatrix}\phi_L(x,y) \\ \chi_L(x,y)\end{pmatrix}
=   \sum_n \begin{pmatrix}\phi^{(n)}_L(x) f_L^n(y) \\ \chi^{(n)}_L(x) g_L^n(y)\end{pmatrix}, 
\label{fermionexpnsn1}
\end{equation}
\begin{equation} 
\Psi_R(x,y) = \begin{pmatrix}\phi_R(x,y) \\ \chi_R(x,y) \end{pmatrix} 
=   \sum_n \begin{pmatrix}\phi^{(n)}_R(x) f_R^n(y) \\ \chi^{(n)}_R(x) g_R^n(y) \end{pmatrix}. 
\label{fermionexpnsn2} 
\end{equation}\\

$\Gamma$-matrices satisfy Clifford algebra in 4$+$1 dimensions:  $\left\{\Gamma_M, \Gamma_N\right\}=2 \mbox{g}_{MN}$, where ($M,N=0,\ldots 4$)\footnote{$M,N =4$ corresponds to the $y$-direction in space.} with the metric  $\mbox{g}_{MN} \equiv {\rm diag}(1,-1,-1,-1,-1)$. The $\Gamma$ matrices are defined as $\Gamma^M \equiv \{\gamma^{\mu}, i\gamma^5$\}, with ($\mu=0,\ldots ,3$).
The 5-D covariant derivative is defined as $D_M\equiv\partial_M+i{g^5_2}\frac{\sigma^{a}}{2}W_M^{a}+i{g^5_1}\frac{Y}{2}B_M$, with ${g^5_2}$ and ${g^5_1}$ are the respective 5-D $SU(2)_L$ and $U(1)_Y$ gauge coupling constants. ${\sigma^{a}}\over 2$ and $\frac Y2$ are the corresponding generators. 

Strength of the BLKT for the fermion fields is parametrised by $r_f$ which we choose to be the same for $\Psi_L$ and $\Psi_R$ to respect the {\em chiral symmetry}. We choose equal strengths of the boundary terms at the fixed points ($y=0$ and $y=\pi R$) which leads to KK-parity\footnote{After restricting the $y$-direction between
$0$ and $\pi R$ and imposing a discrete $Z_2$ symmetry on the action the residual symmetry of the action is a reflection symmetry along 
the line $y = {\pi R \over 2}$. The transformation of KK-excitations under this reflection is characterised by KK-parity.  An immediate consequence of KK-parity conservation is, in any interaction vertex involving KK-excitations, the sum of KK-numbers must be an even integer.} conserving interactions.
We will stick to this convention while defining the gauge and Yukawa interactions in the following. However, in general one can adopt asymmetric
boundary terms leading to non-conservation of KK-parity. Phenomenology of such KK-parity-non-conserving scenario can be found in \cite{ddrs1, lhc}.

One can easily obtain the equation of motion governing the dynamics of $f_{L(R)}$ and $g_{L(R)}$ from the above action. With appropriate boundary conditions, solutions are as the following \cite{carena, flacke, ddrs2}: 
\begin{eqnarray}
f_L^n = g_R^n = N^f_n \left\{ \begin{array}{rl}
                \displaystyle \frac{\cos\left[m_{f^{(n)}} \left (y - \frac{\pi R}{2}\right)\right]}{\cos[ \frac{m_{f^{(n)}} \pi R}{2}]}  &\mbox{for $n$ even,}\\
                \displaystyle \frac{{-}\sin\left[m_{f^{(n)}} \left (y - \frac{\pi R}{2}\right)\right]}{\sin[ \frac{m_{f^{(n)}} \pi R}{2}]} &\mbox{for $n$ odd,}
                \end{array} \right.
                \label{flgr}
\end{eqnarray}
and
\begin{eqnarray}
g_L^n =-f_R^n = N^f_n \left\{ \begin{array}{rl}
                \displaystyle \frac{\sin\left[m_{f^{(n)}} \left (y - \frac{\pi R}{2}\right)\right]}{\cos[ \frac{m_{f^{(n)}} \pi R}{2}]}  &\mbox{for $n$ even,}\\
                \displaystyle \frac{\cos\left[m_{f^{(n)}} \left (y - \frac{\pi R}{2}\right)\right]}{\sin[ \frac{m_{f^{(n)}} \pi R}{2}]} &\mbox{for $n$ odd.}
                \end{array} \right.
\end{eqnarray}
$N^f_n$, being the normalisation for $n^{th}$ KK-mode, could be determined form orthonormality conditions:
\begin{equation}\label{orthonorm}
\begin{aligned}
&\left.\begin{array}{r}
                  \int_0 ^{\pi R}
dy \; \left[1 + r_{f}\{ \delta(y) + \delta(y - \pi R)\}\right]f_L^mf_L^n\\
                  \int_0 ^{\pi R}
dy \; \left[1 + r_{f}\{ \delta(y) + \delta(y - \pi R)\}\right]g_R^mg_R^n
\end{array}\right\}=&&\delta^{n m}~;
&&\left.\begin{array}{l}
                 \int_0 ^{\pi R}
dy \; f_R^mf_R^n\\
                 \int_0 ^{\pi R}
dy \; g_L^mg_L^n
\end{array}\right\}=&&\delta^{n m}~,
\end{aligned}
\end{equation}

and it takes the form:
{\small
\vspace*{-.3cm}
\begin{equation}\label{norm}
N^f_n=\sqrt{\frac{2}{\pi R}}\Bigg[ \frac{1}{\sqrt{1 + \frac{r^2_f m^2_{f^{(n)}}}{4} + \frac{r_f}{\pi R}}}\Bigg].
\end{equation}
}
%Above KK-solutions satisfy following orthonormality conditions: 
$m_{f^{(n)}}$ is the $n^{th}$ KK-mode solution of the following transcendental equations \cite{carena, ddrs2}: 
\begin{eqnarray}
  \frac{r_{f} m_{f^{(n)}}}{2}= \left\{ \begin{array}{rl}
         -\tan \left(\frac{m_{f^{(n)}}\pi R}{2}\right) &\mbox{for $n$ even,}\\
          \cot \left(\frac{m_{f^{(n)}}\pi R}{2}\right) &\mbox{for $n$ odd.}
          \end{array} \right.   
          \label{fermion_mass}      
 \end{eqnarray}

We would like to focus on the other relevant interactions (gauge, scalar and Yukawa) which are also important for our discussion \cite{flacke, zbb}:
\begin{eqnarray}
\label{gauge}
S^W_{gauge} &=& -\frac{1}{4}\int d^5 x \Big[W^{MNi}W_{MN}^{i}+r_V\{ \delta(y) + \delta(y - \pi R)\}W^{\mu \nu i}W_{\mu \nu}^{i} \Big],\\
\label{higgs}
S_{scalar} &=& \int d^5 x  \Big[\left(D^{M}\Phi\right)^{\dagger}\left(D_{M}\Phi\right) + r_{\phi}\{ \delta(y) + \delta(y - \pi R)\}\left(D^{\mu}\Phi\right)^{\dagger}\left(D_{\mu}\Phi\right) \Big], \\
\label{yukawa}
S_{Yukawa} &=& -\int d^5 x  \Big[\lambda^5_t\;\bar{\Psi}_L\widetilde{\Phi}\Psi_R 
  +r_y \;\{ \delta(y) + \delta(y-\pi R) \}\lambda^5_t\bar{\phi_L}\widetilde{\Phi}\chi_R+\textrm{h.c.}\Big].
\end{eqnarray}
 $W_{MN}^i \equiv (\partial_M W_N^i - \partial_N W_M^i-g^5_2f^{ijk}W_M^jW_N^k)$ represents the $SU(2)_L$ ($i=1, 2, 3$) field strength tensor. $\Phi$ is the Higgs doublet and $\widetilde{\Phi}$ satisfy the condition $\widetilde{\Phi}\equiv i\sigma^2 \Phi^\ast$ with $\sigma^2$ being the Pauli matrix. $r_V$  and $r_{\phi}$ are the coefficient of BLKT for gauge and scalar fields respectively while $r_y$ represents the coefficient of boundary terms for Yukawa interactions. $\lambda^5_t$ denote the Yukawa interactions strength for the third generations in 5-D theory. Further, the 5-D gauge coupling $g^5_2$ is related to the 4-D coupling $g_2$ through,
\[g^5_2 = g_2 ~\sqrt{\pi R \left(1+\frac{r_V}{\pi R}\right)}.\]

Appropriate KK-expansion of the fields which are involved in the above actions can be schematically written as:
\begin{equation}\label{W}
W^i_{\mu}(x,y)=\sum_n W_{\mu}^{i(n)}(x) a^n(y)~;~~
W^i_{4}(x,y)=\sum_n W_{4}^{i(n)}(x) b^n(y),
\end{equation}
and
\begin{equation}\label{phi}
\Phi(x,y)=\sum_n \Phi^{(n)}(x) h^n(y).
\end{equation}

Note, that $y$-dependent profiles for the $W^\mu$ and $W^4$ must be different in the view of the fact that $W^\mu$ KK-tower must have a $0$-mode, while the later shouldn't have such a mode in its KK-tower.
In general, due to electroweak symmetry breaking, the eigenvalue equation for the  gauge boson contains a term proportional to ($r_{\phi} - r_V $)\cite{gf}. Consequently, KK-solutions of the gauge bosons are in general, different form Eq.\;\ref{flgr}. To avoid  unnecessary complication we set $r_{\phi} = r_V$, which we will keep in the rest of our analysis. Consequently, gauge boson KK-excitations have masses $m_{V^{(n)}}$ ($ = m_{\phi^{(n)}}$)  which satisfy the same transcendental equations given in Eq.\;\ref{fermion_mass} with $r_f$
replaced by $r_V$ ($= r_{\phi}$). Furthermore, this assumption helps us to fix the gauge properly in this non-trivial scenario. For a detail discussion of the $W^\mu$ and $W^4$ $y$-dependent profiles we refer to the ref.\;\cite{gf}.

One must supplement the action for fermions and gauge bosons with  the gauge-fixing action \cite{gf}:

\begin{eqnarray}
\label{gauge_fix}
S_{\rm gf}^{W} &=& -\frac{1}{\xi _y}\int d^5x\Big\vert\partial_{\mu}W^{\mu +}+\xi_{y}(\partial_{y}W^{4+}+iM_{W}\phi^{+}\{1 + r_{V}\left( \delta(y) + \delta(y - \pi R)\right)\})\Big \vert ^2 .\nonumber \\
\end{eqnarray}
Here, $W^{4+}$ is the $5^{th}$ component of charged $SU(2)$  gauge boson, while $\phi^+$ is the $T_3 = \frac{1}{2}$ component of the Higgs doublet. 
$M_W$ is the 0-mode $W$-boson mass and $\xi _y$ is connected with {\em physical} gauge fixing parameter $\xi$ (with values 0 (Landau gauge), 1 (Feynman gauge) or $\infty$ (Unitary gauge)) via, 
\begin{equation}
\xi= \xi_y \{1 + r_{V}\left( \delta(y) + \delta(y - \pi R)\right)\}.
\end{equation}

Using  Eqs.\;\ref{gauge}, \ref{higgs},  and Eq.\;\ref{gauge_fix} we can write the bi-linear terms involving the KK-modes of $W^{4(n)\pm}$ and $\phi^{(n)\pm}$ in $R_{\xi}$ gauge as:

\begin{equation}
\label{gauge_mix}      
\mathcal{L}_{W^{4(n)\pm}  \phi^{(n)\pm}} = - \begin{pmatrix}
W^{4(n)-} & \phi^{(n)-}
\end{pmatrix}
\begin{pmatrix}
M_{W}^{2}+\xi m^2_{V^{(n)}} & -i(1-\xi)M_{W}m_{V^{(n)}} \\ i(1-\xi)M_{W}m_{V^{(n)}} & m^2_{V^{(n)}}+\xi M_{W}^{2}
\end{pmatrix}
\begin{pmatrix}
W^{4(n)+} \\ \phi^{(n)+}
\end{pmatrix}.
\end{equation}
In addison to Goldstone bosons (with $n^{th}$ mode mass square $\xi  (M_W^2+m^2_{V^{(n)}}) $):
 $$
G^{(n)\pm} = \frac{\left(m_{V^{(n)}}W^{4(n)\pm}\pm  iM_{W}\phi^{(n)\pm}\right)}{M_{W^{(n)}}},
$$
we have additional physical charged Higgs pair (with $n^{th}$ mode mass square ($M_W^2+m^2_{V^{(n)}}$)):
$$
H^{(n)\pm} = \frac{\left(m_{V^{(n)}}\phi^{(n)\pm}\pm iM_{W}W^{4(n)\pm}\right)}{M_{W^{(n)}}}.
$$
The fields $W^{\mu (n)\pm}$ and $H^{(n)\pm}$ have the same mass eigenvalue $M_{W^{(n)}} \equiv
\sqrt {M_W^2+m^2_{V^{(n)}}}$ and in 't-Hooft Feynman gauge ($\xi  = 1$), $G^{(n)\pm}$ also correspond to the same mass eigenvalue.

Let us now examine the mixing of the quark sector. 
This mixing is only important for top quarks as it is driven by the Yukawa coupling.
Using  the modal expansions for fermions given in Eqs.\;\ref{fermionexpnsn1} and \ref{fermionexpnsn2} and substituting these
 in the actions given in Eq.\;\ref{factn} and Eq.\;\ref{yukawa} we can find the bi-linear terms involving  the doublet and singlet states of the quarks. The mass matrix for $n^{th}$ KK-level is as the following: 

\begin{equation}
\label{fermion_mix}
-\begin{pmatrix}
\bar{\phi_L}^{(n)} & \bar{\phi_R}^{(n)}
\end{pmatrix}
\begin{pmatrix}
m_{f^{(n)}}\delta^{nm} & m_{t} {\mathscr{I}}^{nm}_1 \\ m_{t} {\mathscr{I}}^{mn}_2& -m_{f^{(n)}}\delta^{mn}
\end{pmatrix}
\begin{pmatrix}
\chi^{(m)}_L \\ \chi^{(m)}_R
\end{pmatrix}+{\rm h.c.},
\end{equation}
where $m_t$ is the top quark mass and $m_{f^{(n)}}$ are the solutions of transcendental equations given in Eq.\;\ref{fermion_mass}. The overlap integrals (${\mathscr{I}}^{nm}_1$ and ${\mathscr{I}}^{nm}_2$) are of the form:
  \[ {\mathscr{I}}^{nm}_1=\left(\frac{1+\frac{r_f}{\pi R}}{1+\frac{r_y}{\pi R}}\right)\times\int_0 ^{\pi R}\;dy\;
\left[ 1+ r_y \{\delta(y) + \delta(y - \pi R)\} \right] g_{R}^m f_{L}^n,\] \;\;{\rm and}\;\;\[{\mathscr{I}}^{nm}_2=\left(\frac{1+\frac{r_f}{\pi R}}{1+\frac{r_y}{\pi R}}\right)\times\int_0 ^{\pi R}\;dy\;
 g_{L}^m f_{R}^n .\]

For both the cases of $n=m$ and $n\neq m$ the integral ${\mathscr{I}}^{nm}_1$ is non zero. But for $r_y = r_f$, this integral equal to 1 (when $n =m$) or 0 ($n \neq m$). And the integral ${\mathscr{I}}^{nm}_2$ is non zero only when  $n =m$ and equal to 1 in the limit $r_y = r_f$. Using this equality (fermion and Yukawa
BLT) condition we can easily avoid the mode mixing and construct simpler form of the fermion mixing matrix. In the rest of our analysis we will stick to the choice of equal $r_y$ and $r_f$. 

The resulting matrix (given in Eq.\;\ref{fermion_mix}) can be diagonalised by following bi-unitary transformations for the left- and right-handed fields respectively:
\begin{equation}
U_{L}^{(n)}=\begin{pmatrix}
\cos\alpha_{tn} & \sin\alpha_{tn} \\ -\sin\alpha_{tn} & \cos\alpha_{tn}
\end{pmatrix},~~U_{R}^{(n)}=\begin{pmatrix}
\cos\alpha_{tn} & \sin\alpha_{tn} \\ \sin\alpha_{tn} & -\cos\alpha_{tn}
\end{pmatrix},
\end{equation}
where $\alpha_{tn}[ = \frac12\tan^{-1}\left(\frac{m_{t}}{m_{f^{(n)}}}\right)]$ is the mixing angle. The gauge eigen states $\Psi_L(x,y)$ and $\Psi_R(x,y)$ and mass eigen states $T^1_t$ and $T^2_t$ are related by the following relations:
%
%\begin{equation}
%\begin{eqnarray}
\begin{tabular}{p{8cm}p{8cm}}
{\begin{align}
&{\phi^{(n)}_L} =  \cos\alpha_{tn}T^{1(n)}_{tL}-\sin\alpha_{tn}T^{2(n)}_{tL},\nonumber \\
&{\chi^{(n)}_L} =  \cos\alpha_{tn}T^{1(n)}_{tR}+\sin\alpha_{tn}T^{2(n)}_{tR},\nonumber
\end{align}}
%\end{eqnarray}
&
%\begin{eqnarray}
{\begin{align}
&{\phi^{(n)}_R} =  \sin\alpha_{tn}T^{1(n)}_{tL}+\cos\alpha_{tn}T^{2(n)}_{tL},\nonumber \\
&{\chi^{(n)}_R} =  \sin\alpha_{tn}T^{1(n)}_{tR}-\cos\alpha_{tn}T^{2(n)}_{tR}.
%\end{eqnarray}
\end{align}}
\end{tabular}
%\end{equation}
%

The eigen states are degenerate in mass, values of which are given by: $m_{T^{1(n)}_t}=m_{T^{2(n)}_t}=\sqrt{m_{t}^{2}+m^2_{f^{(n)}}} \equiv M_{t^{(n)}}$.

Feynman rules (in 't-Hooft Feynman gauge) which follow from the above action and necessary for our calculation are listed in Appendix C. 
One can see that some of the interactions listed in Appendix C have been modified with respect to their UED counterparts by  some overlap integrals. These overlap integrals are generated while arriving at the 4-D effective action from 5-D one by inserting the appropriate $y$-dependent profiles in the (5-D) action and  integrating over the $y$-direction. Here we present the  overlap integrals which will frequently occur in our calculation. 

\begin{itemize}
 \item Interaction of a $0$-mode fermion with a KK-fermion ($n^{th}$ mode) and a KK-gauge boson ($m^{th}$ mode):
\begin{eqnarray}
\label{gauge_overlap1}
I^{n m}_V=\sqrt{\pi R\left(1+\frac{r_V}{\pi R}\right)}\times\left\{\begin{array}{rl}
                \displaystyle \int_0 ^{\pi R}
dy \; \left[1 + r_{f}\{ \delta(y) + \delta(y - \pi R)\}\right]a^mf_L^nf_L^0~,\\
                \displaystyle \int_0 ^{\pi R}
dy \; \left[1 + r_{f}\{ \delta(y) + \delta(y - \pi R)\}\right]a^mg_R^ng_R^0~.
\end{array}\right.
\label{gff}
\end{eqnarray}

\item Interaction of a $0$-mode fermion with a KK-fermion ($n^{th}$ mode) and  the $m^{th}$ KK-mode of $5^{th}$ component of a gauge boson:
\begin{eqnarray}
\label{gauge_overlap2}
I^{\prime n m}_V=\sqrt{\pi R\left(1+\frac{r_V}{\pi R}\right)}\times\left\{\begin{array}{rl}
                \displaystyle \int_0 ^{\pi R}
dy \; b^mg_L^nf_L^0~,\\
                \displaystyle \int_0 ^{\pi R}
dy \; b^mf_R^ng_R^0~.
\end{array}\right.
\label{aff}
\end{eqnarray}
In Eq.\;\ref{gff} $a^m$ is the wave function ($m^{th}$ KK-mode) for gauge field, with Lorentz index $\mu$.  While in Eq.\;\ref{aff} $b^m$ is the wave function ($m^{th}$ KK-mode) for $5^{th}$ component of the corresponding gauge field.\\

\item Yukawa interactions of fermions and scalars:
\begin{eqnarray}
\label{youkawa_overlap}
I^{n m}_Y=\sqrt{\pi R\left(1+\frac{r_V}{\pi R}\right)}\times\left\{\begin{array}{rl}
                \displaystyle \int_0 ^{\pi R}
dy \; \left[1 + r_{f}\{ \delta(y) + \delta(y - \pi R)\}\right]h^mf_L^ng_R^0~,\\
                \displaystyle \int_0 ^{\pi R}
dy \; \left[1 + r_{f}\{ \delta(y) + \delta(y - \pi R)\}\right]h^mg_R^nf_L^0~.
\end{array}\right.
\end{eqnarray}
In the above, $h^m$ is the wave function ($m^{th}$ KK-mode) for scalar field.\\
\end{itemize}

Now for, $r_{\phi}=r_V$ we have $a^m\equiv h^m$  and for $r_f=r_y$, $I^{n m}_V\equiv I^{n m}_Y$, for $n=m$ let us call it $I^n_1$, and is given by:  
\begin{equation}
I^n_1 = 2\sqrt{\frac{1+\frac{r_V}{\pi R}}{1+\frac{r_f}{\pi R}}}\left[ \frac{1}{\sqrt{1 + \frac{r^2_f m^2_{f^{(n)}}}{4} + \frac{r_f}{\pi R}}}\right]\left[ \frac{1}{\sqrt{1 + \frac{r^2_V m^2_{V^{(n)}}}{4} + \frac{r_V}{\pi R}}}\right]\frac{m^2_{V^{(n)}}}{\left(m^2_{V^{(n)}} - m^2_{f^{(n)}}\right)}\frac{\left(r_{f} - r_{V}\right)}{\pi R},
\label{i1}
\end{equation}

and we denote $I^{\prime n m}_V=I^{nm}_2$. Let us call it $I^n_2$ for $n=m$ and is given by: 
\begin{equation}
I^n_2 = 2\sqrt{\frac{1+\frac{r_V}{\pi R}}{1+\frac{r_f}{\pi R}}}\left[ \frac{1}{\sqrt{1 + \frac{r^2_f m^2_{f^{(n)}}}{4} + \frac{r_f}{\pi R}}}\right]\left[ \frac{1}{\sqrt{1 + \frac{r^2_V m^2_{V^{(n)}}}{4} + \frac{r_V}{\pi R}}}\right]\frac{m_{V^{(n)}}m_{f^{(n)}}}{\left(m^2_{V^{(n)}} - m^2_{f^{(n)}}\right)}\frac{\left(r_{f} - r_{V}\right)}{\pi R}.
\label{i2}
\end{equation}

\begin{figure}[t]
\begin{center}
\vspace*{-.3cm}
\includegraphics[scale=.99,angle=0]{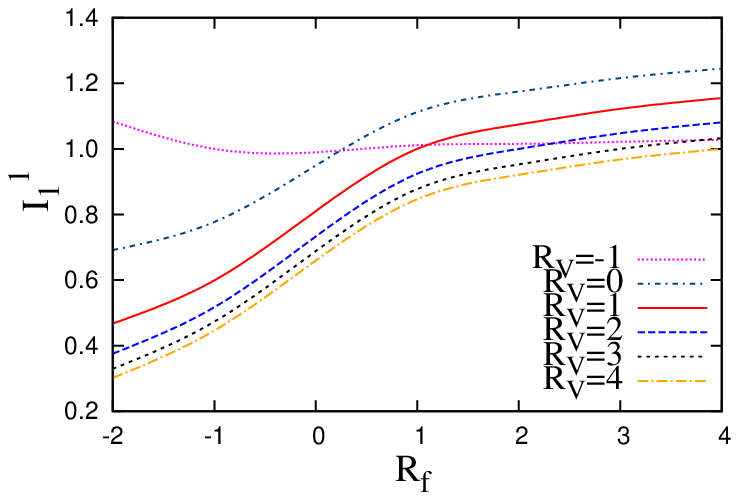}
\includegraphics[scale=.99,angle=0]{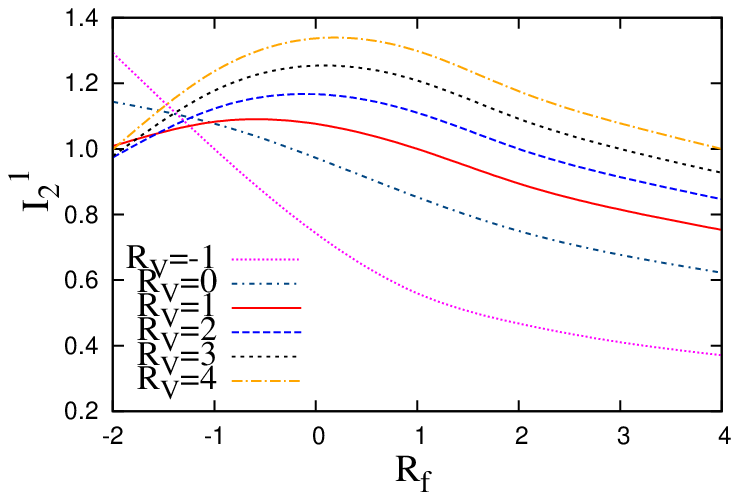}
\caption{Variation of the overlap integrals $I^1_1$ (left panel) and $I^1_2$ (right panel) defined in Eqs.\;\ref{i1} and \ref{i2} with $R_f$ for six different values of $R_V$.}
\label{int}
\end{center}
\end{figure}

These two overlap integrals play a crucial role in our analysis. Many of the couplings used in the following calculation are hallmarked by the presence these integrals which can be explicitly seen in the Feynman rules listed in Appendix C. First of all, in the limit $r_f = r_V$, 
both of these overlap integrals become unity, setting all the nmUED couplings equal to their UED values. For the purpose of illustration, 
 variations of the integrals $I^1_1$ and $I^1_2$ ($n=m=1$) with $r_f/R(=R_f)$ and $r_V/R(=R_V)$ have been presented in Fig.\;\ref{int}. Left (right) panel shows the variation of $I^1_1$ ($I^1_2$) with $R_f$ for six different values of $R_V$. We have presented these integrals for both positive as well as negative values of $R_V$ and $R_f$. Let us first discuss the pattern of these integrals for positive values of $R_V$. In this case, while $I^1_1$ increases with $R_f$, $I^1_2$ initially increases before it starts decreasing with $R_f$.   On the other hand if $R_V$ is negative,  $I^1_1$ is almost independent of $R_f$, but $I^1_2$ steadily decreases with $R_f$ over the entire range of its values used in Fig.\;\ref{int}. At this point we would like to make some comments which could help the reader to understand the nature of the plots. The first two factors (one under the square root sign and the other in the first square bracket involving $r_f$ only) decreases with $R_f$ in both of these integrals.  Among the remaining factors, term proportional to $m^2_{V^{(1)}}$ (in $I^1_1$)  slowly diminish in magnitude with $R_f$. The remaining factor proportional to $(R_f - R_V)$ increases with $R_f$. When $R_V$ is negative ($= -1$), due to an interplay between the  term proportional to $m^2_{V^{(1)}}$ and remaining factors, $I^1_1$ remains almost independent with $R_f$. This explains the 
 qualitative departure of the overlap integral $I^1_1$ for negative and positive values of $R_V$.  Furthermore, as $m_{f^{(1)}}$ is  a decreasing function of $R_f$, in $I^1_2$ the presence of a factor proportional to $m_{V^{(1)}} m_{f^{(1)}}$ (instead of $m^2_{V^{(1)}}$ in $I^1_1$) offers additional damping to
 the magnitude of $I^1_2$.  And apart from some lower values of $R_f$, $I^1_2$ always decreases with $R_f$.

Before going into the main calculation, let us comment on the range of values of BLT parameters used in our analysis. In general BLT parameters can positive or negative. 
A careful look into the Eq.\;\ref{norm} would reveal that,  for ${r_f}/{R}=-\pi$ the 0-mode solution becomes divergent. And beyond  ${r_f}/{R} = - \pi$ the 0-mode fields become ghost-like.  
Any other values of BLT parameters greater than $- \pi$ are acceptable. However, as  BLT parameters change from positive to negative domain, corresponding KK-masses would increase 
thus diminishing the values of loop
functions described below. This in turn decrease the magnitudes of the branching ratios of our interest. 

\vspace*{-.2cm}
\section{\boldmath{$B_{s(d)}\rightarrow\mu^{+}\mu^{-}$} in nmUED}

\begin{figure}[H]
\begin{center}
\includegraphics[scale=0.6,angle=0]{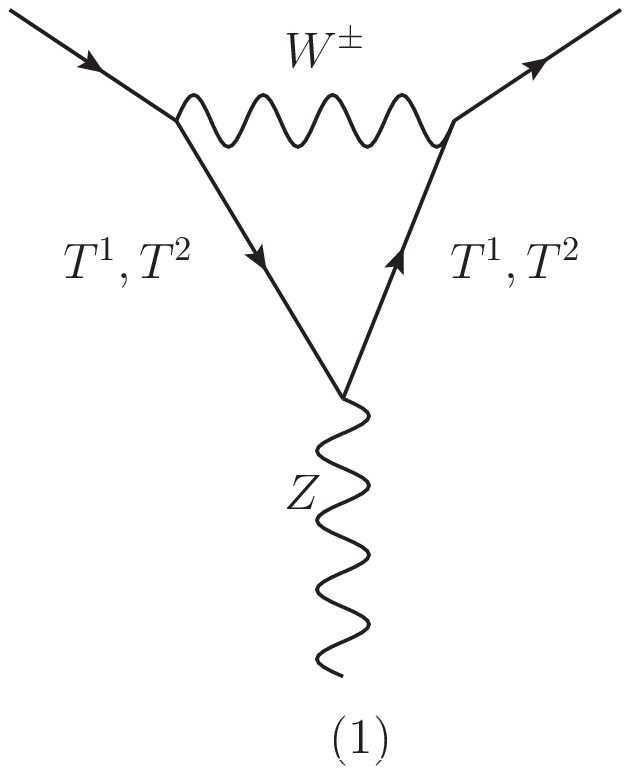}
\includegraphics[scale=0.6,angle=0]{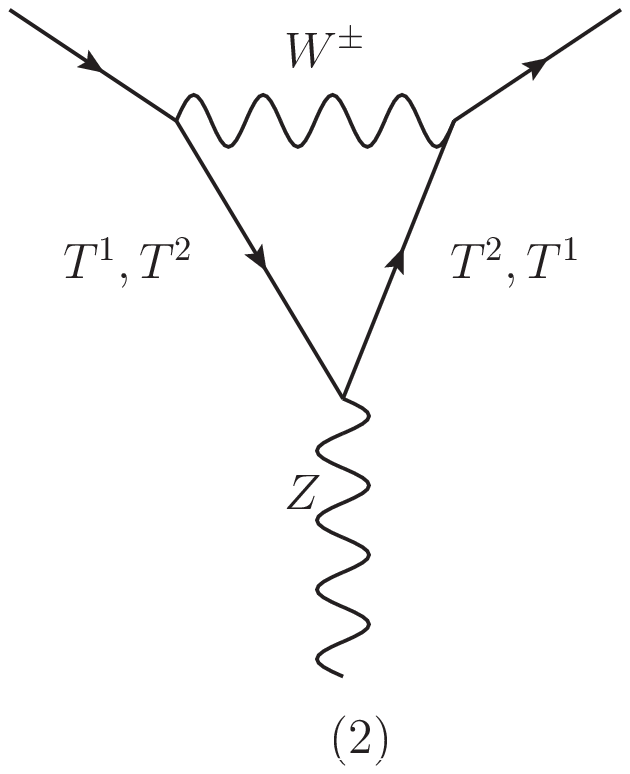}
\includegraphics[scale=0.6,angle=0]{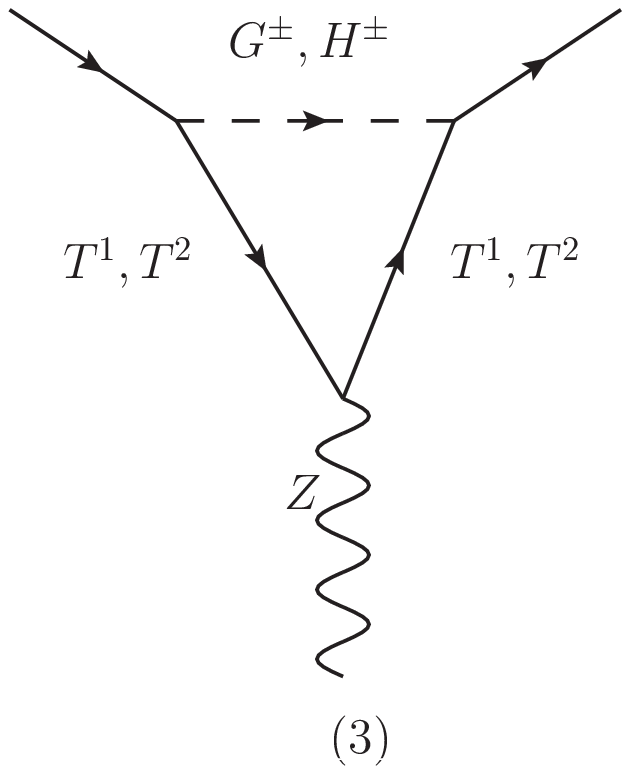}
\includegraphics[scale=0.6,angle=0]{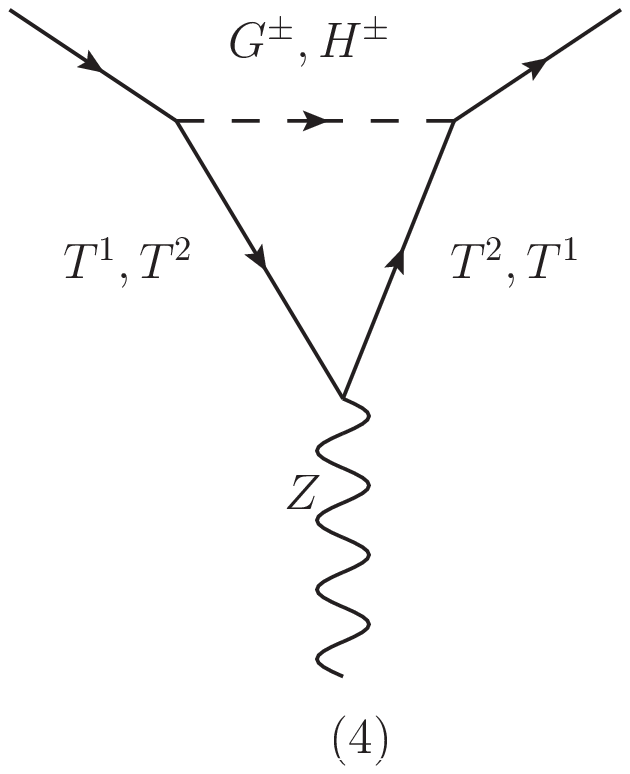}
\includegraphics[scale=0.6,angle=0]{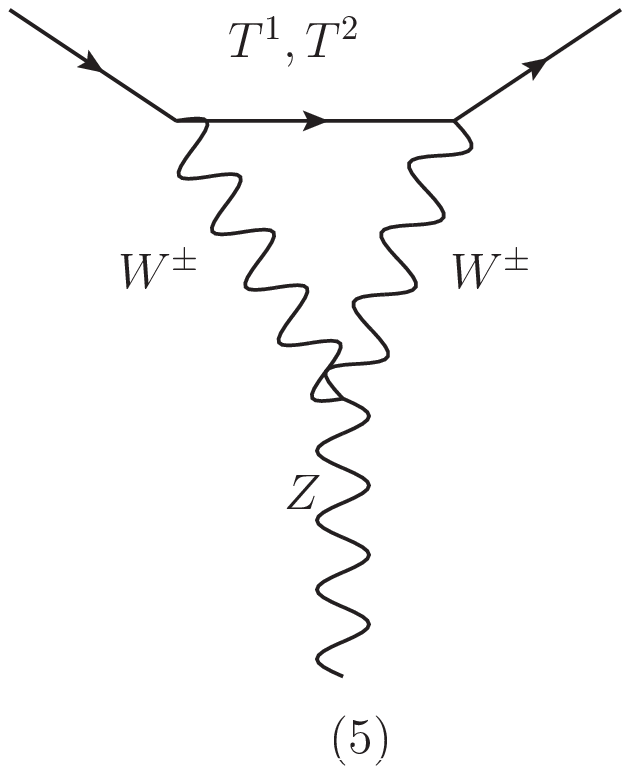}
\includegraphics[scale=0.6,angle=0]{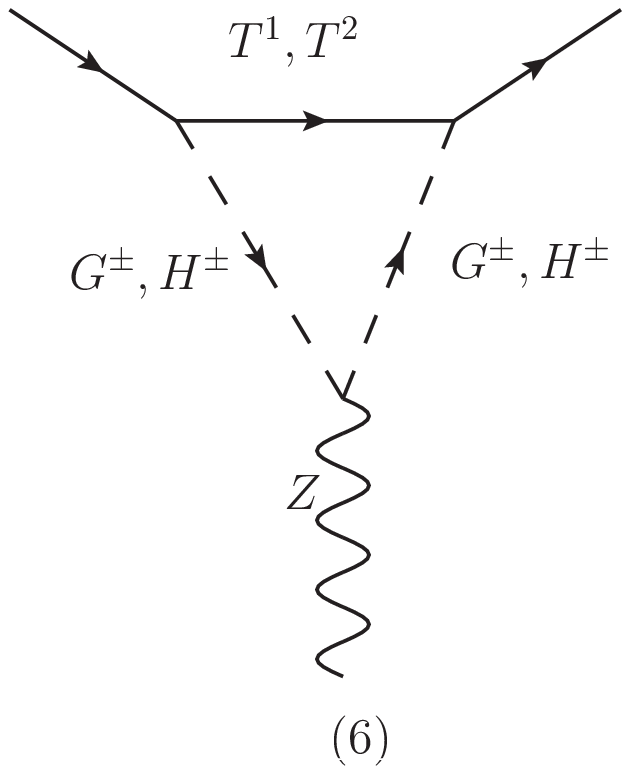}
\includegraphics[scale=0.6,angle=0]{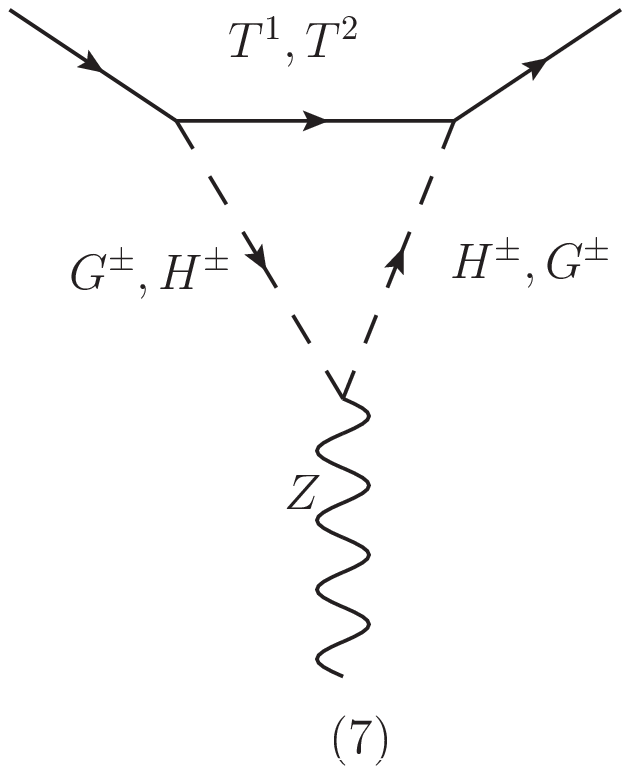}
\includegraphics[scale=0.6,angle=0]{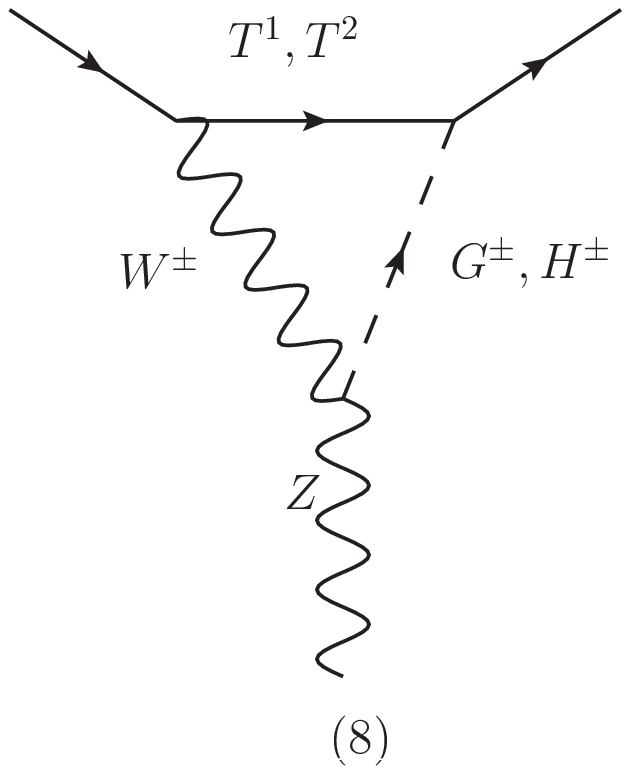}
\caption{$Z$-penguin diagrams contributing to the decay of $B_{s(d)}$.}
\label{pen}
\end{center}
\end{figure}

\begin{figure}[H]
\begin{center}
\includegraphics[scale=0.6,angle=0]{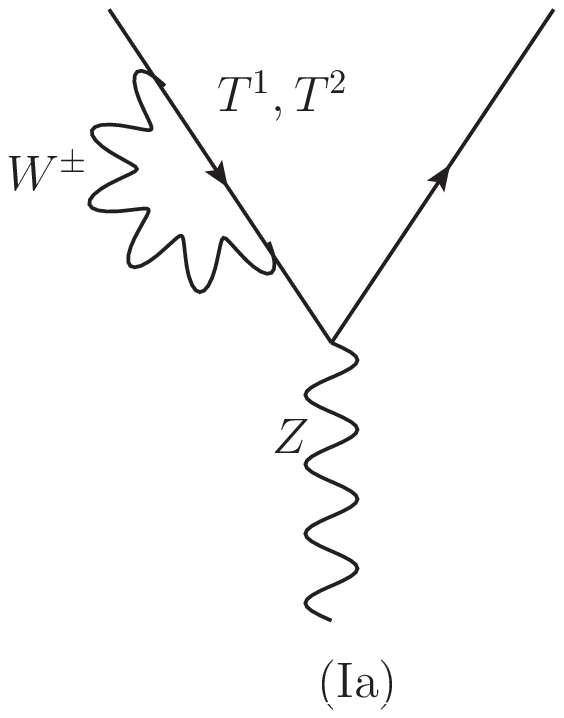}
\includegraphics[scale=0.6,angle=0]{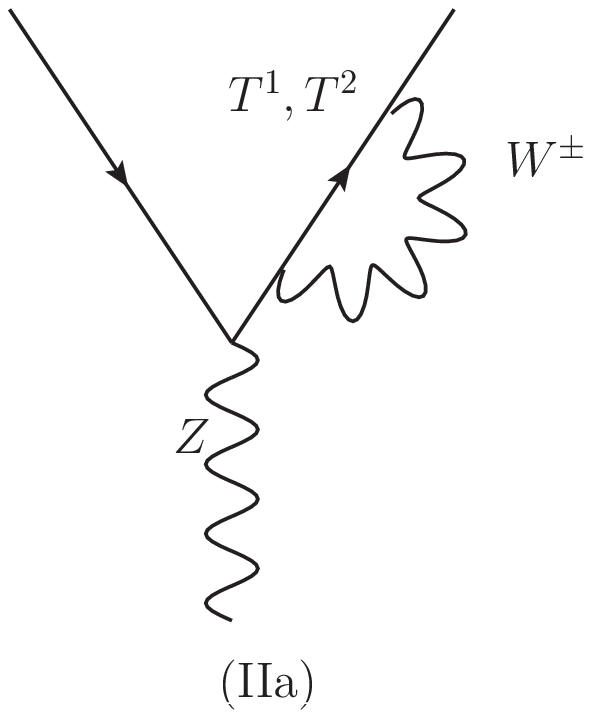}
\includegraphics[scale=0.6,angle=0]{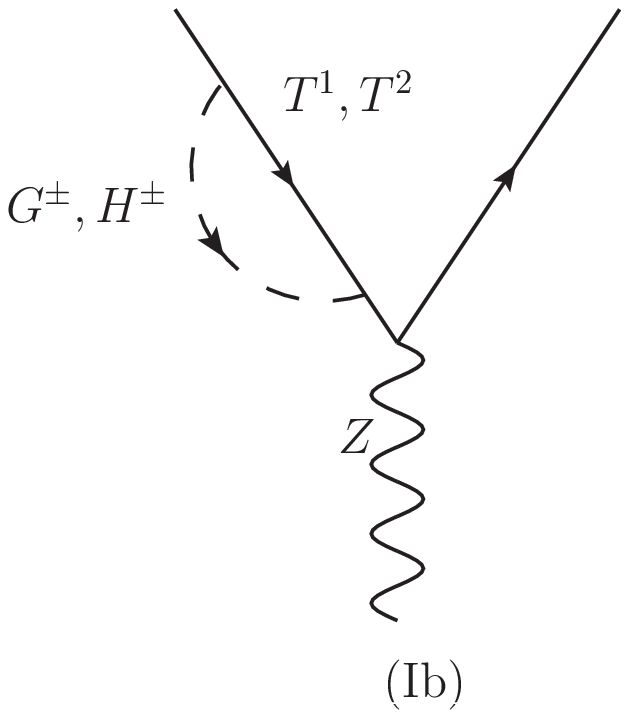}
\includegraphics[scale=0.6,angle=0]{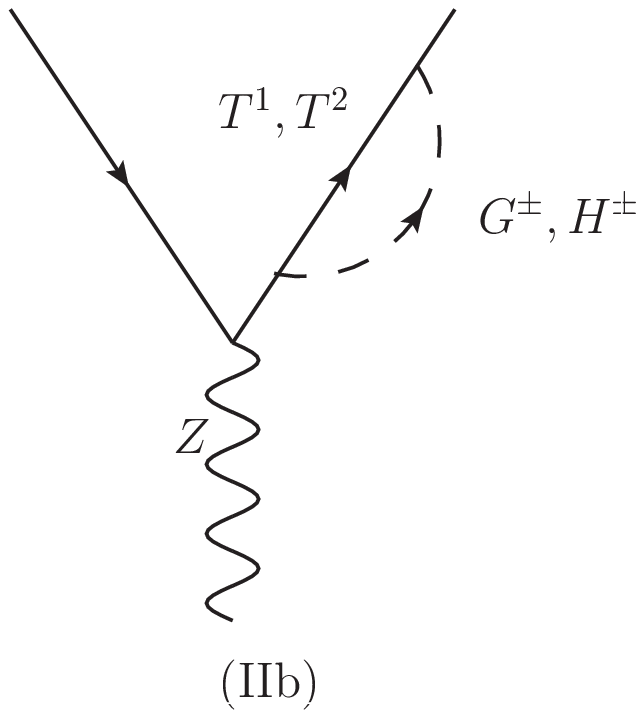}
\caption{Self-energy diagrams contributing to the decay of $B_{s(d)}$.}
\label{self}
\end{center}
\end{figure}

Stage  has now been set to discuss the subject of our attention, namely the branching ratio of $B_{s(d)}$ into $\mu^+ \mu^-$ pair. 
The effective Hamiltonian for {$B_{s(d)}\rightarrow\mu^{+}\mu^{-}$} decay 
is given by:
\begin{equation}\label{hyll}
{\cal H}_{\rm eff} = -\frac{G_F}{\sqrt{2}}\frac{\alpha} 
{2\pi \sin^2 \theta_{w}} V^\ast_{tb} V_{ts(d)}
\eta_Y Y (x_t, r_f, r_V, R^{-1}) \left[\bar b \gamma_\mu \, (1 - \gamma_5) s(d)\right]\; \left[ \bar \mu \gamma^\mu \, (1 - \gamma_5)\mu \right]+ {\rm h.c.}~,   
\end{equation}
where $G_F$ is the Fermi constant, $V_{ij}$ are the Cabibbo-Kobayashi-Maskawa (CKM) matrix elements, $\alpha$ is the fine structure constant, $\eta_Y$ is the QCD factor and $\theta_w$ is the Weinberg angle. The function $Y(x_t, r_f, r_V, R^{-1})$ is the total contributions coming from $Z$-penguins, self-energy and box diagrams:
\begin{equation}\label{yyx}
Y(x_t, r_f, r_V, R^{-1})  =C(x_t, r_f, r_V, R^{-1})+B(x_t, r_f, r_V, R^{-1}),
\end{equation}

where $C(x_t, r_f, r_V, R^{-1})$ originates from penguins and self energy diagrams (Fig.\;\ref{pen}  and Fig.\;\ref{self} respectively) and is defined as:
\begin{equation}\label{cn}
C(x_t, r_f, r_V, R^{-1})=C_0(x_t)+
\sum_{n=1}^\infty C_n(x_{t(n)},x_{u(n)}).
\end{equation}
Here, $C_0(x_t)$ is the SM contribution:
\begin{equation}
C_0(x_t)={\frac{x_t}{8}}\left[\frac{x_t-6}{x_t-1}+\frac{3x_t+2}
{(x_t-1)^2}\;\ln x_t\right],
\label{c0-eq}
\end{equation}
while the second term represents the total KK-contribution that is being
calculated from the $Z$-penguin diagrams and self-energy diagrams (Figs.\;\ref{pen} and \ref{self}) originating from the nmUED framework. The function $C_n(x_{t(n)},x_{u(n)})$ is defined as: 
\begin{equation}
C_n(x_{t(n)},x_{u(n)})= F(x_{t(n)})-F(x_{u(n)}),
\label{subtract}
\end{equation}
where $F(x_{t(n)})$ and $F(x_{u(n)})$ arises respectively, from the contributions of the 
$T^{1(n)}_{t}$, $T^{2(n)}_{t}$ and $T^{1(n)}_{u}$, $T^{2(n)}_{u}$ modes. The function $F(x_{t(n)})$ takes the form:
\begin{equation}
 F(x_{t(n)}) = \sum_{i=1}^8 F_i(x_{t(n)}) + \left(\frac12 - \frac13
\sin^2 \theta_w \right) \sum_{i=1}^2 \Delta S_i ( x_{t(n)}).
\end{equation}
 Moreover $F_i$ denote the contributions those are coming from 
 $Z$-penguin diagrams (1-8 in Fig.\;\ref{pen}) and $\Delta S_i$ represent the contributions of self-energy diagrams (Ia-IIb in Fig.\;\ref{self}) which are necessary for calculation of the electroweak counter terms. While the function $F(x_{u(n)})$ is given by:
\begin{equation}
F(x_{u(n)})=F(x_{t(n)})\bigg|_{x_{t}\rightarrow 0}.
\end{equation}
$F(x_{u(n)})$ are subtracted in the above (Eq.\;\ref{subtract}) to take into account of the contributions of KK-excitation of first two generations of quarks 
in $Z$-penguins and self-energy diagrams, exploiting the GIM-mechanism. $F(x_{t(n)})$ in nmUED framework is given by the following expression which is drastically different from that of UED due to the presence of integrals $I^n_1$and $I^n_2$:
%\vspace*{-.5cm}
{\small
\begin{eqnarray}\label{fn}
\hspace*{-2cm}
&& F(x_{t(n)})=\left[ \frac{1}{8}\left\{-\ln{M^2_{t^{(n)}}}-\frac{3}{2}+h_q\left(x_{t(n)}\right)\right\}-\frac{1}{4} x_{t(n)} h_q\left(x_{t(n)}\right)\frac{m^2_{f^{(n)}}}{M^2_{t^{(n)}}}\right](I^n_1)^2
\nonumber \\ &&
-\frac{3}{4}\left[-\ln{M^2_{W^{(n)}}}-\frac{1}{6}-x_{t(n)}h_w\left(x_{t(n)}\right)\right](I^n_1)^2
%\nonumber \\ &&
-\frac{1}{2}h_w\left(x_{t(n)}\right)\frac{m^2_{f^{(n)}}}{M^2_{W^{(n)}}}(I^n_1)^2
\nonumber \\ &&
+\bigg[\frac{1}{16}\left\{-\ln{M^2_{t^{(n)}}}-\frac{1}{2}+h_q\left(x_{t(n)}\right)\right\}(I^n_2)^2
%\nonumber \\ &&
-\frac{1}{8}x_{t(n)}h_q\left(x_{t(n)}\right)\left\{\frac{m^4_{f^{(n)}}}{M^2_{t^{(n)}}m^2_{V^{(n)}}}+\frac{m^4_t}{M^2_{t^{(n)}}M^2_W}\right\}(I^n_1)^2\bigg]
\nonumber \\ &&
-\frac{1}{16}\left[2(I^n_2)^2+x_t(I^n_1)^2\right]\left[-\ln{M^2_{W^{(n)}}}+\frac{1}{2}-x_{t(n)}h_w\left(x_{t(n)}\right)\right]
\nonumber \\ &&
+\frac18 \left[
    -\frac{1}{2} \left\{\frac{1+x_{t(n)}}{1-x_{t(n)}} +  \frac{2
    x_{t(n)}^2 \ln x_{t(n)} }{(1-x_{t(n)})^2} \right\} - 
    \ln {M^2_{W^{(n)}}} \right](I^n_1)^2
\nonumber \\ && 
+\frac{1}{16} \left[
   (I^n_2)^2 + x_t(I^n_1)^2 \right]  \left[\frac{1}{2} 
\left\{ \frac{1-3x_{t(n)}}{1-x_{t(n)}} -
          \frac{2 x_{t(n)}^2\ln x_{t(n)} }{(1-x_{t(n)})^2} \right\}   - 
    \ln {M^2_{W^{(n)}}} \right].
\end{eqnarray}
}
Expressions for $h_q$ and $h_w$ are listed in Appendix A (Eqs. A-9 and A-10).

\begin{figure}[H]
\begin{center}
\includegraphics[scale=0.8,angle=0]{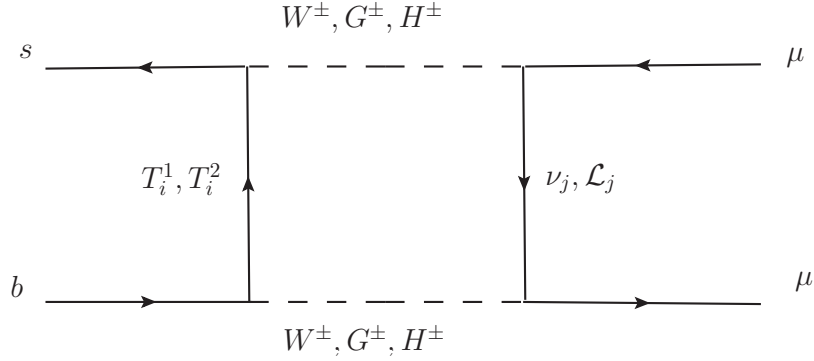}
\caption{Box diagrams contributing to the decay of $B_{s(d)}$.}
\label{box}
\end{center}
\end{figure}

$B(x_t, r_f, r_V, R^{-1})$ in Eq.\;\ref{yyx} is given by:
\begin{equation}\label{bn}
B(x_t, r_f, r_V, R^{-1})=-B_0(x_t)+
\sum_{n=1}^\infty B_n(x_{t(n)}, x_{u(n)}, x_{\nu(n)}).
\end{equation}
$B_0(x_t)$ again represents the SM contribution to the box diagrams:

\begin{equation}
B_0(x_t)=\frac{1}{4}\left[\frac{x_t}{1-x_t}+\frac{x_t}
{(x_t-1)^2}\;\ln x_t\right],
\label{b0-eq}
\end{equation}

and the second term denotes the total KK-contribution that is
evaluated from the box diagrams (Fig.\;\ref{box}). $B_n(x_{t(n)}, x_{u(n)}, x_{\nu(n)})$ is defined as:
\begin{equation}\label{b_n}
B_n(x_{t(n)}, x_{u(n)}, x_{\nu(n)})= H(x_{t(n)},x_{\nu(n)})-H(x_{u(n)},x_{\nu(n)}).
\end{equation}
In nmUED framework, expression for $H(x_{t(n)},x_{\nu(n)})$ is given as the following:

{\small
\begin{eqnarray}\label{hn}
\hspace*{-2cm}
&&H(x_{t(n)},x_{\nu(n)})=(I^n_1)^4\Bigg[ -\frac14 \frac{M_W^2}{M^2_{W^{(n)}}}~ U(x_{t(n)}, x_{\nu(n)})
%\nonumber \\ &&
+\frac12 \frac{M_W^2 M^2_{t^{(n)}}m^2_{f^{(n)}}}{M^6_{W^{(n)}}}~ \widetilde{U}(x_{t(n)}, x_{\nu(n)})
\nonumber \\ &&
+\frac12 \frac{M_W^2 m^2_{f^{(n)}}}{M^6_{W^{(n)}}}\left(\frac{M^2_W m^2_{f^{(n)}}}{m^2_{V^{(n)}}}-m^2_t\right)~ \widetilde{U}(x_{t(n)}, x_{\nu(n)})
%\nonumber \\ &&
-\frac18 \frac{M_W^2 m^2_{f^{(n)}}}{M^6_{W^{(n)}}}\left(\frac{M^2_W m^2_{f^{(n)}}}{m^2_{V^{(n)}}}-m^2_t\right)~ U(x_{t(n)}, x_{\nu(n)})
\nonumber \\ &&
-\frac{1}{16} \frac{M_W^2 M^2_{t^{(n)}}m^2_{f^{(n)}}}{M^6_{W^{(n)}}}~U(x_{t(n)}, x_{\nu(n)})
%\nonumber \\ &&
-\frac{1}{16} \frac{M_W^2}{M^6_{W^{(n)}}}\left(\frac{M^4_Wm^4_{f^{(n)}}}{m^4_{V^{(n)}}}+m^2_tm^2_{f^{(n)}}\right)~U(x_{t(n)}, x_{\nu(n)})\Bigg],
\end{eqnarray}
}
and 
$H(x_{u(n)},x_{\nu(n)})$ is defined through:
\begin{equation}
H(x_{u(n)},x_{\nu(n)})=H(x_{t(n)},x_{\nu(n)})\bigg|_{x_{t}\rightarrow 0}.
\end{equation}

The expressions for the functions $U$ and $\widetilde U$ can be found in Appendix B (Eqs. B-7 and B-8).

The definition of $B_n$ (given in Eq.\;\ref{b_n}) are again  in order to take into account the contributions from KK-excitations of 
first two generations of quarks into the calculation.
%Here,
%\begin{equation}
%H(x_{t(n)},x_{\nu(n)})=H_{WW(n)}+H_{WG(n)}+H_{WH(n)}+H_{GH(n)}+H_{GG(n)}+H_{HH(n)},
%\end{equation}

Note that $x_t=\frac{m^2_t}{M^2_W}, \;{\rm and}\;x_{t(n)}=\frac{M^2_{t^{(n)}}}{M^2_{W^{(n)}}}=\frac{m^2_t+m^2_{f^{(n)}}}{M^2_W+m^2_{V^{(n)}}}$\;, $m_{f^{(n)}}$ and $m_{V^{(n)}}$ are the solutions of Eq.\;\ref{fermion_mass} corresponding to the BLKT parameters $r_f$ and $r_V$ respectively.

Finally one can express the branching ratio of $B_{s(d)}\to \mu^{+}\mu^{-}$ in terms of $Y$ and other parameters as:
\begin{eqnarray}
Br(B_{s(d)}\to \mu^{+}\mu^{-})&=&\tau(B_{s(d)})\frac{G^2_{\rm F}}{\pi}
\left(\frac{\alpha}{4\pi\sin^2\theta_{w}}\right)^2 F^2_{B_{s(d)}}m^2_{\mu} m_{B_{s(d)}} \nonumber \\
&&\sqrt{1-4\frac{m^2_{\mu}}{m^2_{B_{s(d)}}}} |V^\ast_{tb}V_{t{s(d)}}|^2 \;Y^2(x_t, r_f, r_V, R^{-1}), \nonumber \\
\label{br}
\end{eqnarray}

where, $F_{B_{s(d)}}$ is the ${B_{s(d)}}$ meson decay constant and $\tau(B_{s(d)})$ is the life time. $m_\mu$ and $m_{B_{s(d)}}$ are the masses of muon and $B_{s(d)}$ meson respectively.

We would like to point out that the total contribution from UED is obtained by summing $C_n$s and $B_n$s over KK-modes (starting from $n = 1$). It can be checked easily that these quantities 
tend to vanish when $R^{-1} \rightarrow \infty$, showing the decoupling nature of the new physics in this case. 
 While estimating the contribution from the KK-excitation, in the one loop mediated amplitudes, we have  
used only those interactions which couples a 0-mode field to pair of KK-excitations with same KK-number. However, in a manifestly KK-parity conserving theory, there can be non zero interactions among 
a 0-mode field with two different KK-excitations, say, with unequal KK-number $n$ and $m$, modulo $n +m$ is an even integer.  We have explicitly checked that $0-n-m$ interactions are suppressed with respect to the 
$0-n-n$ interactions \cite{zbb} and we neglect those while estimating the $B_{s(d)}\rightarrow \mu^+ \mu^-$ interactions.
 
In our calculation, we have neglected the masses of the external quarks and while evaluating 
the $Z$-penguins and box diagrams, momenta of the external legs have been assumed to be vanishing.

\section{Results}
We are now equipped to present our results. Before we proceed further  let us  present the experimentally measured value \cite{ntr} of $Br(B_{s(d)}\rightarrow \mu^+ \mu^-$) along with the SM prediction 
\cite{sm, ntr} for the same in Table \ref{t:4}. 
Central values of experimentally measured branching ratios  of $B_s$ ($B_d$) are nearly one (two)-standard deviation away from its SM prediction.

\begin{table}[H]
\begin{center}
\begin{tabular}{|c|c|c|}
\hline
Branching ratio& SM prediction & Experimental value \\ \hline
$B_{s}\to \mu^{+}\mu^{-}$ & $(3.66\pm0.23)\times10^{-9}$ & $(2.8^{+0.7}_{-0.6})\times10^{-9}$ \\ \hline
$B_{d}\to \mu^{+}\mu^{-}$ & $(1.06\pm0.09)\times10^{-10}$ & $(3.9^{+1.6}_{-1.4})\times10^{-10}$ \\ \hline
\end{tabular}
\caption{SM prediction for $B_{s(d)} \rightarrow \mu^+ \mu^-$ branching ratios and their experimental measured values.}
\label{t:4}
\end{center}
\end{table}

We will now discuss the method, that has been used in the following, for the numerical estimation of $Br(B_{s}\to \mu^{+}\mu^{-})$.  We will follow similar procedure for the estimation of  $Br(B_{d}\to \mu^{+}\mu^{-})$. Numerical value of the $B_s$ branching ratio, quoted in Table\;\ref{t:4} includes higher order
corrections (NNLO-QCD and NLO-EW corrections \cite{sm}) to the branching ratio in the framework of the SM.  The total decay amplitude in the UED framework is a sum of contributions 
from different KK-levels (including the contribution from $0$-level which is nothing  but the SM). We expect that higher order corrections to
the decay amplitude originating from higher KK-modes would be in the same ballpark that of the SM ($0$-mode). So to (approximately) 
take into account the effects of higher order corrections into our result we have estimated $Br(B_{s}\to \mu^{+}\mu^{-})$ in UED as,\footnote{This method is somehow similar to find a $k$-factor 
(the ratio of branching ratios with higher order correction and without any correction in our case) in the SM  and then multiply the leading order branching ratio in UED by this $k$-factor. In absence of a
full higher order correction in UED framework, this approximation is not very bad, as the higher KK-modes of $t$-quark
and $W$-bosons, running in the triangle or box diagrams, have similar couplings to the SM particles in the initial and the final state.} 
\begin{equation}  Br(B_{s}\to \mu^{+}\mu^{-})_{UED} =  Br(B_{s}\to \mu^{+}\mu^{-})_{SM} \; \left( \frac{Y} {Y_0} \right)^2.
\label{mtd} 
\end{equation}
Here, $Y$ total contribution (originating from $0$ and higher KK-modes) defined in Eq.\;\ref{yyx}, while $Y_0 (\equiv C_0 + B_0)$ is
only the $0$-mode (SM) contribution. Expressions for $C_0$ and $B_0$ are given in Eq.\;\ref{c0-eq} and Eq.\;\ref{b0-eq} respectively. For 
numerical estimation of the above branching ratio, we have used the central value ($3.66  \times10^{-9}$) for $Br(B_{s}\to \mu^{+}\mu^{-})_{SM}$, following the ref.\;\cite{ntr}.

In the above, we have expressed $Br(B_{s}\to \mu^{+}\mu^{-})_{UED}$ in terms of $Br(B_{s}\to \mu^{+}\mu^{-})_{SM}$. 
Consequently, the relevant input parameters for numerical evaluation of the above branching ratio are top-quark and $W$-boson masses
respectively, which are necessary for evaluation of the functions $Y_0$ and $Y$. We use the values of these parameters following the ref.\;\cite{pdg}. Values of these parameters used in ref.\;\cite{sm} are same as the values in ref.\;\cite{pdg}\footnote{We have used $M_W=80.38$ GeV and  $m_t=173.21$ GeV.}.

Any contribution to the $B_{s(d)} \rightarrow \mu^+ \mu^-$ branching ratio from the SM and beyond the SM goes into the function $Y$ as defined in Eq.\;\ref{br}.
The dominant contribution to $Y$ comes from the diagrams (3), (4), (6) and (7) of Fig.\;\ref{pen} ($Z$-penguins); (Ib) and (IIb) of Fig.\;\ref{self} (self-energies). These amplitudes are proportional to $g_2 \lambda_t^2$, where $\lambda_t$ being the top-Yukawa coupling. Box diagram contributions are not significant in comparison to the penguins \footnote{As for example, in nmUED model, for $R^{-1} = 1$ TeV, $R_f = R_V = 1$ after summing over 5 KK-levels 
$C = 0.9116$ while $B = 0.1856$. For $R^{-1} = 2$ TeV, $R_f = R_V = 1$, $C = 0.8915$ while $B = 0.1846$. The smallness of the box diagram contribution in comparison to the penguins, has also been pointed out in ref.\;\cite{buras}}.  One must perform a sum over the KK-modes while 
evaluating the $C$ and $B$-functions, which contribute to $Y$. In view of a recent analysis correlating the SM Higgs mass and cut-off scale of UED model \cite{kksum}, we restrict ourselves upto 5 KK-levels. Previous practice was to use 20-30 KK-levels while summing up the contributions from KK-modes. However, we have explicitly checked that numerical results would not change to a significant level as  the sum over the KK-modes, in this case is converging\footnote{It has been shown in \cite{gg_converge}, the KK-sum in case of one loop calculation (like the one in the present case) in UED with one extra space like dimension is always converging.}. At the end of following sub-section, we will present a table comparing the $B_s$ branching ratios calculated with 5 KK-levels vis-a-vis 20 KK-levels in support of our assumption.

The main results of our analysis {\em i.e.} variation of the branching ratio ($B_{s}\rightarrow \mu^{+}\mu^{-}$) with $R^{-1}$ for several values of $R_f$ are 
presented in Fig.\;\ref{bs}. Same have been presented for $B_d$ in Fig.\;\ref{bd}. Six different panels show the dependence of branching ratios with $R_V$. We will be discussing the implications of these results 
in the framework of nmUED. But before delving into that, we would like to discuss the consequences of the result in the framework of UED itself.

\subsection{Reviewing the limits on \boldmath{$R^{-1}$} in UED framework}
\begin{figure}%[t]
%\vspace*{-2cm}
\begin{center}
\includegraphics[scale=1.1,angle=0]{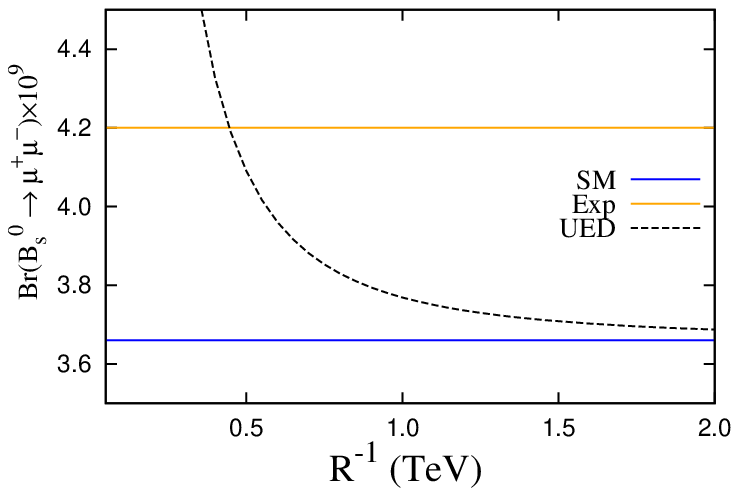}
\includegraphics[scale=1.1,angle=0]{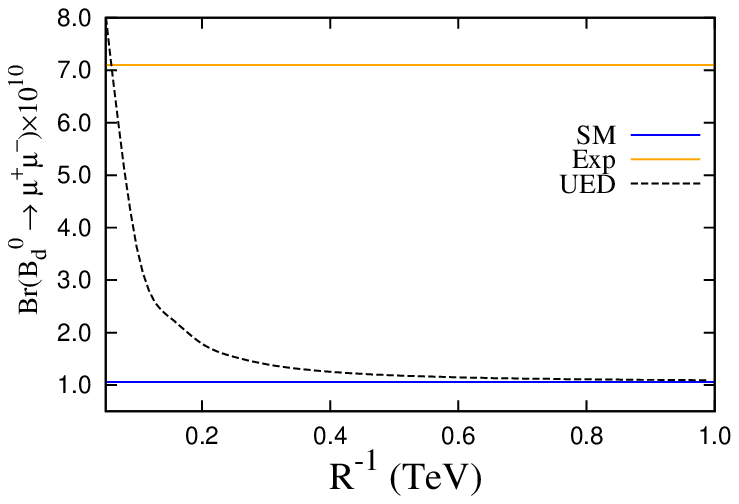}
\caption{Left (right) panel: Variation of the branching ratio ($B_{s}(B_{d})\rightarrow \mu^{+}\mu^{-}$) with $R^{-1}$ in UED model.  The black dashed line in each panel represents the UED curve (corresponds to 
$r_V/R=R_V=0$ and $r_f/R=R_f=0$ case). The dark shaded horizontal line represents the SM prediction for the above branching ratio, while the 
light shaded horizontal line represents the 95 \% C.L. upper limit on the experimentally measured $B_{s(d)}$ branching ratio to muon pair.}
\label{UED}
\end{center}
\end{figure}

To start with we will focus on the $B_{s(d)}$ meson branching ratio to $\mu^+ \mu^-$ in the UED frame work. The expression for $F^n$ and $H^n$ defined in Eqs.\;\ref{fn} and \ref{hn} would result into their UED 
forms once we set $r_V = r_f =0$. In this limit the overlap integrals $I^n_1$ and $I^n_2$ become unity and the masses of the KK-excitations in $n^{th}$ KK-mode become equal to  $nR^{-1}$. We check 
 that our expressions of $C$ and $B$ in this limit agree exactly with the one in ref.\;\cite{buras}\footnote{The authors of ref.\;\cite{buras} have not considered any radiative corrections to the KK-masses, consequently mass of a $n^{th}$ KK-excitation is $nR^{-1}$ in their analysis.}. In Fig.\;\ref{UED} we have plotted the $B_{s(d)} \rightarrow \mu^+ \mu^-$ branching ratio with $R^{-1}$ in UED framework (black dashed curve in each panel).  Steady decrement of  the branching ratio with increasing $R^{-1}$ is an artefact of increasing KK-masses with $R^{-1}$. This line intersects the horizontal line (light shaded) ($4.2 \times 10^{-9}$ for $B_s$ and $7.1 \times 10^{-10}$ for $B_d$) which correspond to the 95\% C.L. upper limit  on the experimentally measured branching ratio of $B_s$ (left panel of Fig.\;\ref{UED}) and $B_d$ (right panel of Fig.\;\ref{UED}). Thus the $R^{-1}$ values corresponding to these intersections would give rise to a 95\% C.L. lower limit on the value of $R^{-1}$. One can see this limit from $B_s$ decay comes out to be 454 GeV.  While deriving this limit, we have neglected the theoretical error  on $Br(B_s \rightarrow \mu^+ \mu^-$). One can easily see from Table\;\ref{t:4}, that theoretical error on the SM estimate of this branching ratio is around 6\% and we expect that theory error in the UED framework will be of the same order.  And this is small compared to the  experimental error (nearly 25 \%) on the measured value of the branching ratio. Our results presented in Fig.\;\ref{UED}, thus correspond to a vanishing theoretical error on the $B_s$ branching ratio. 
 
The lower bound on $R^{-1}$ which results from $B_d$ decay ($R^{-1}\sim 60$ GeV) is not so promising. To put our observation into the proper context we would briefly mention the bounds on $R^{-1}$ that result from other processes. Consideration of  $(g-2)_\mu$ \cite{mu2}, $\rho$-parameter \cite{rho}, FCNC process \cite{buras, buras3, fcnc} and electroweak observables like $R_b$ \cite {zbb, elctrowk} would result into a lower limit on $R^{-1}$ which is in the ballpark of 300 GeV.  Consideration of 
 radiative $B$-meson decay puts a lower limit of  600 GeV on $R^{-1}$ \cite{bsg}. While the projected lower limit on $R^{-1}$ using  tri-lepton signal at 8 TeV LHC is 1.2 TeV  \cite{belayev}.
 
\subsection{Bounds from the nmUED}
\begin{figure}%[t]
%\vspace*{-2cm}
\begin{center}
\includegraphics[scale=1.1,angle=0]{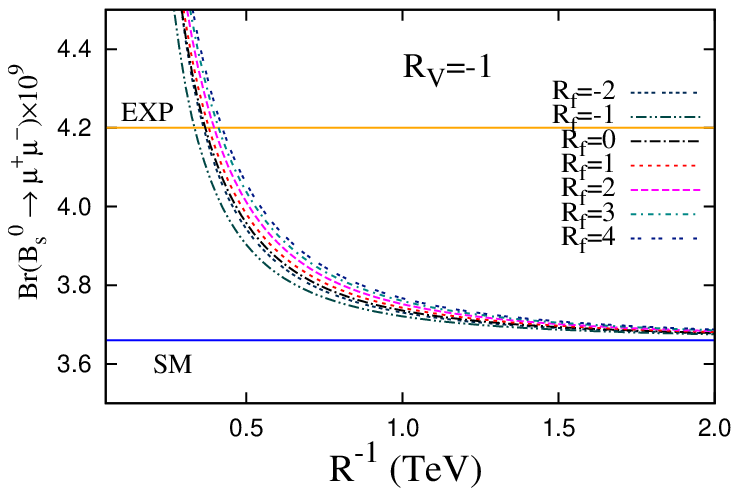}
\includegraphics[scale=1.1,angle=0]{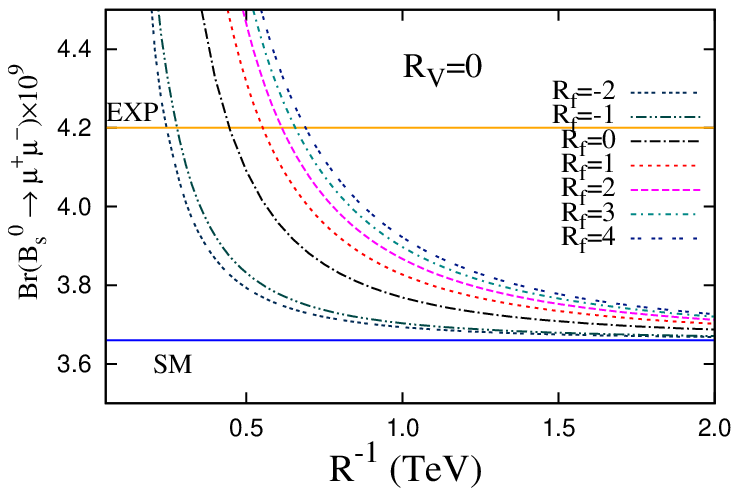}
\includegraphics[scale=1.1,angle=0]{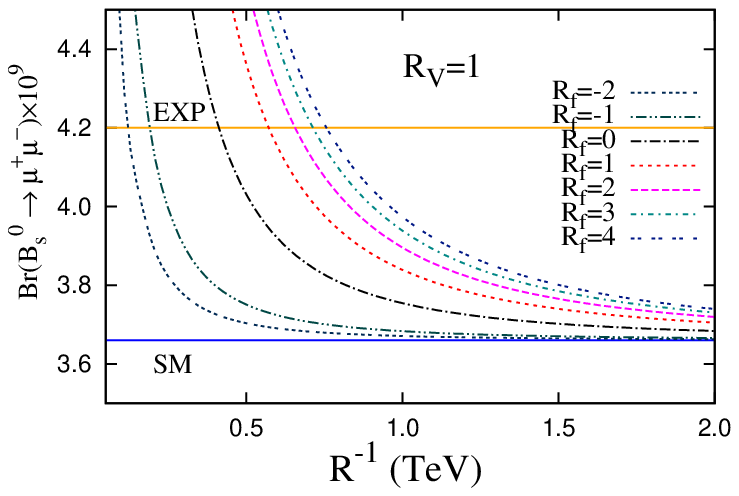}
\includegraphics[scale=1.1,angle=0]{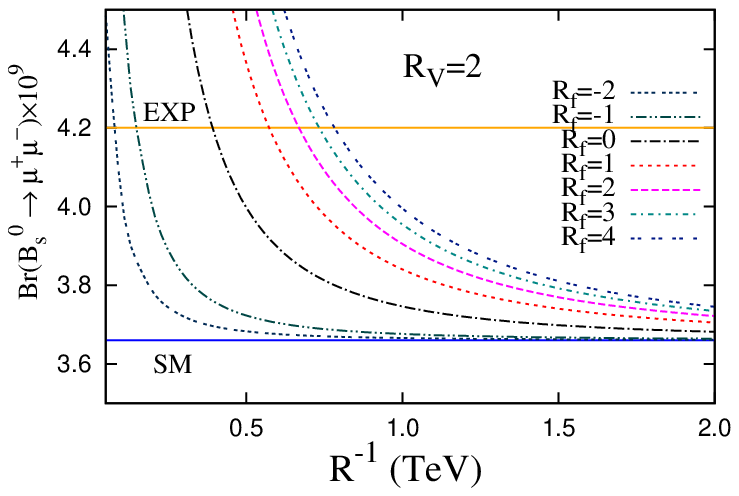}
\includegraphics[scale=1.1,angle=0]{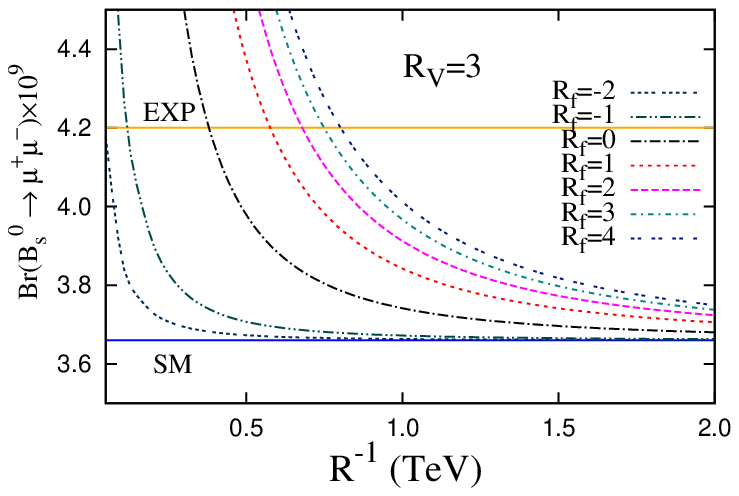}
\includegraphics[scale=1.1,angle=0]{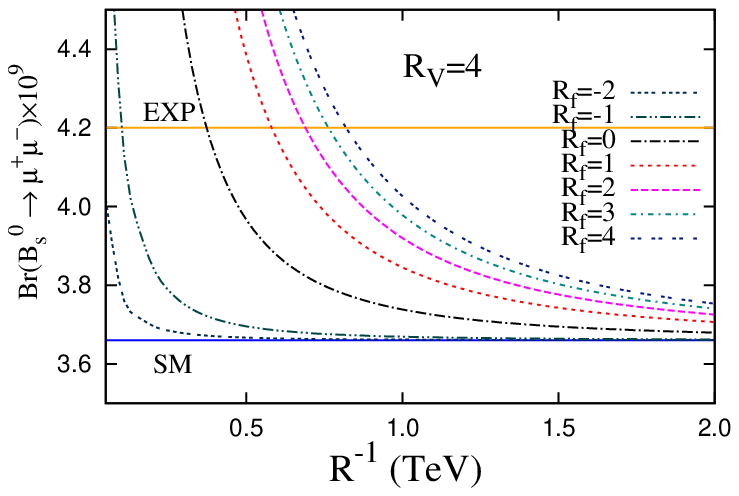}
\caption{Variation of the branching ratio ($B_{s}\rightarrow \mu^{+}\mu^{-}$) with $R^{-1}$ for several values of $R_f=r_f/R$ . The six panels correspond to different $R_V=r_V/R$. The dark shaded horizontal line represents the SM prediction for the above branching ratio, while the 
light shaded horizontal line represents the 95 \% C.L. upper limit on the experimentally measured $B_s$ branching ratio to muon pair.}
\label{bs}
\end{center}
\end{figure}

\begin{figure}%[t]%[H]
\begin{center}
%\vspace*{-2cm}
%\includegraphics[width=5.5cm,height=5.5cm, angle =0]{Br_d_rg1_big}
%includegraphics[width=5.5cm,height=5.5cm, angle =0]{Br_d_rg2_big}
%includegraphics[width=5.5cm,height=5.5cm, angle =0]{Br_d_rg3_big}
\includegraphics[scale=1.1,angle=0]{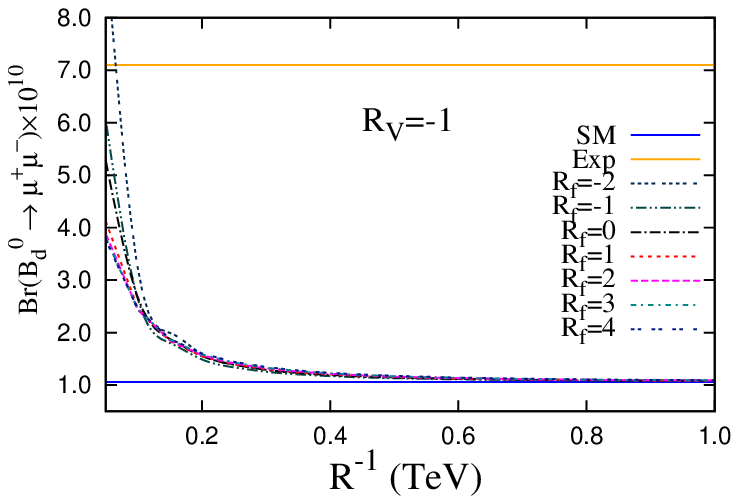}
\includegraphics[scale=1.1,angle=0]{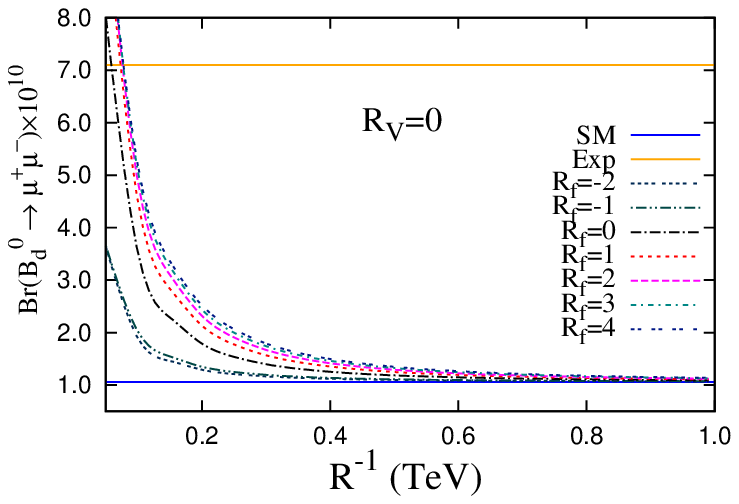}
\includegraphics[scale=1.1,angle=0]{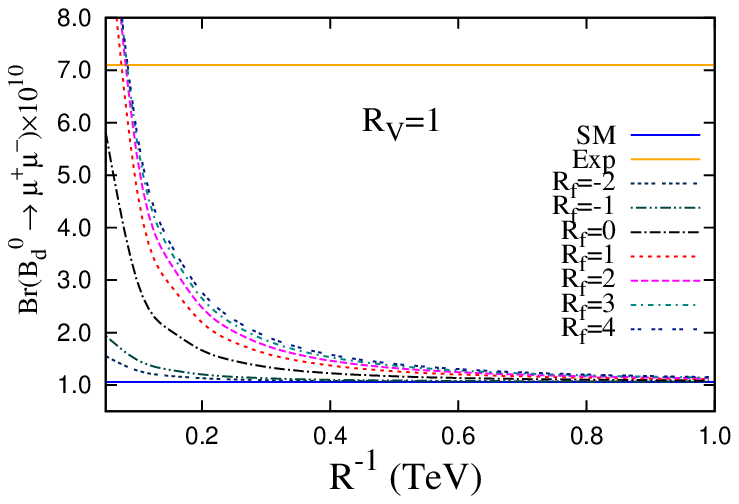}
\includegraphics[scale=1.1,angle=0]{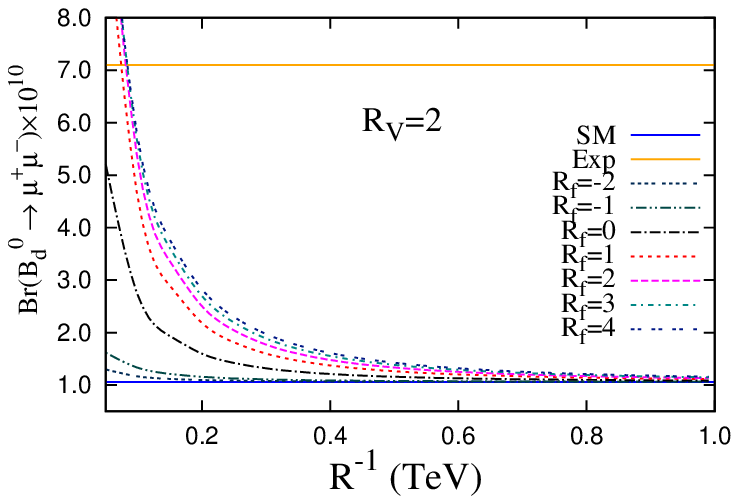}
\includegraphics[scale=1.1,angle=0]{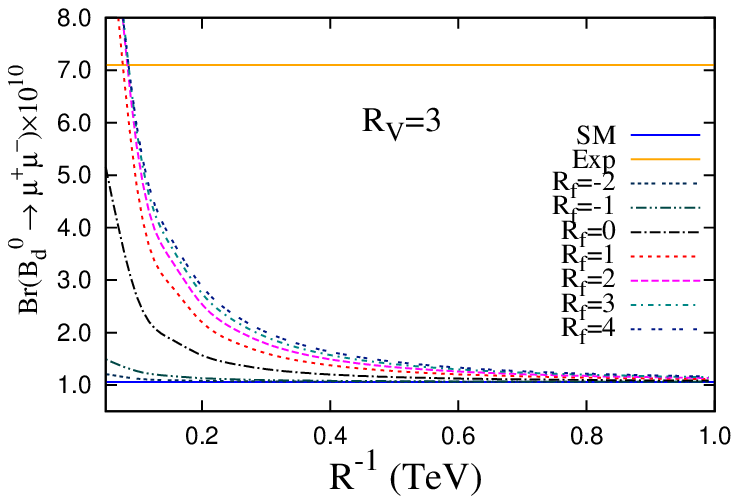}
\includegraphics[scale=1.1,angle=0]{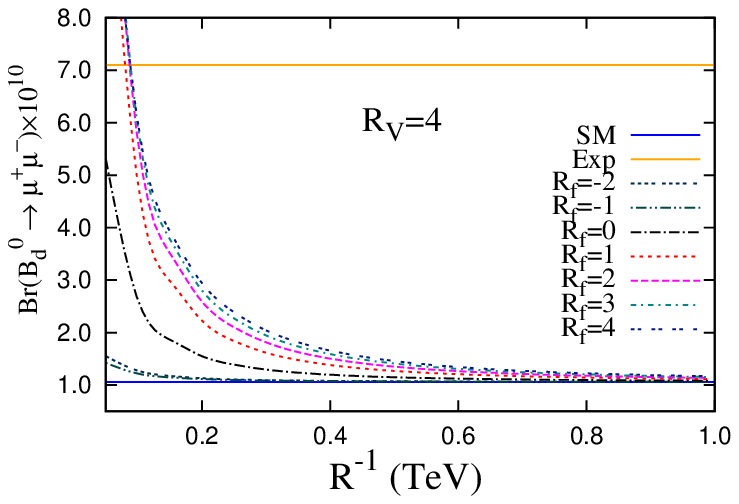}
\caption{Variation of the branching ratio ($B_{d}\rightarrow \mu^{+}\mu^{-}$) with $R^{-1}$ for several values of $R_f=r_f/R$ . The six panels correspond to different $R_V=r_V/R$. The dark shaded horizontal line represents the SM prediction for the above branching ratio, while the 
light shaded horizontal line represents the 95 \% C.L. upper limit on the experimentally measured $B_d$ branching ratio to muon pair.}
\label{bd}
\end{center}
\end{figure}

We are now in a position to discuss our results in the context of nmUED. Numerical results from the calculation presented in the previous section 
are presented in Figs.\;\ref{bs} and \ref{bd} for $B_s$ and $B_d$ meson branching ratios to muon pair respectively. In each of the figures 
six different panels represents the results for six different choices for gauge BLKT, $R_V$. While in each panel different line 
represents the variation of the branching ratio with $R^{-1}$ with different values  of fermion BLKT coefficients, $R_f$.

The monotonic decreasing nature of the branching ratio with increasing $R^{-1}$ is the same as in UED and has been explained in the above. The dependence on the other parameters could be  easily understood as the following. For a given $R^{-1}$, masses of KK-excitations of 
fermions, gauge bosons and scalars decrease with increasing $R_f$ and $R_V$. This in turn, increase the decay width of $B_s$ and $B_d$
to $\mu^+ \mu^-$. Apart from KK-masses the other determining  factors in our calculation are the overlap integrals defined in Eqs.\;\ref{i1} and \ref{i2}.  We have already noted their dependencies on $R_f$ and $R_V$.  

One can see that $B_s \rightarrow \mu^+ \mu^-$ branching ratio
increases with the increment of both of these BLKT coefficients. Dependence on $R_V$ is mild while it is more sensitive to any change of 
$R_f$. This can be understood again by looking at the interactions used in this calculation listed in Appendix C.  A careful look at the Feynman rules reveals that $I^n_1$ and $I^n_2$ modify the interactions of $3^{rd}$ generations of quarks with charged-Higgs scalar and 
$5^{th}$ component of $W$-boson respectively. The first one being top-Yukawa, is dominant over the second one, which is $SU(2)$
gauge interaction. Thus $I^n_1$ which appears with top-Yukawa, has a better control on the $B_s \rightarrow \mu^+ \mu^-$ amplitude, which 
increases both with $R_f$ and $R_V$. Dependence  of $Br(B_d \rightarrow \mu^+ \mu^-)$ on $R^{-1}$ and BLKT parameters can be explained in the same way. The decay of $B_d$ to $\mu^+ \mu^-$ is suppressed with respect to the $B_s$ by a factor of $\left( \frac{V_{td}}{V_{ts}}\right)^2$. Finally we would like to add that for negative values of BLKT parameters, above branching ratio is always less than the value for the same in the UED model. This could be easily accounted by the suppression of decay amplitude due to  heavier (than in the UED case) KK-excitation in nmUED framework for negative values of BLKT parameters.    

Finally in Fig.\;\ref{cont} (right panel), the exclusion region in $R_f - R^{-1}$ plane, has been presented for six different choices of $R_V$. The region 
under a particular line (corresponding to a fixed value of $R_V$) has been excluded at 95\% C.L. by comparing the experimentally measured $Br(B_s \rightarrow \mu^+ \mu^-)$ to its theoretical prediction in the framework of nmUED. This plot is a summary of the results presented in 
Fig.\;\ref{bs}.  The lines represent the contours of constant branching ratios of $B_s \rightarrow \mu^+ \mu^-$ corresponding to the 95\% C.L. upper limit ($4.2 \times 10^{-9}$) of its experimentally determined value. Features of this contours can be easily understood in the light of our discussion above. Higher values of $R^{-1}$ would increase the KK-masses hence pulling down the decay width (and branching ratio). To compensate this one must increase $R_f$ and $R_V$, which control the decay dynamics in two ways. First of all, these would pull down the masses and $R_f$ would also increase the couplings through the overlap integral $I^n_1$. While an increasing $R_V$ would decrease $I^n_1$ but increase
$I^n_2$. Overall, with increasing $R_V$, decay width mildly increase in contrast to $R_f$, which would increase the decay width more sharply with its increment.
 
We would like to show the difference in the derived lower limits on $R^{-1}$ due to taking into account 5 KK-levels vis-a-vis 20 KK-level while performing the KK-level summation. One can 
see from Table\;\ref{sum-kkn}, that the lower limits on $R^{-1}$ (in TeV) derived from our analysis is not very much sensitive to the number of KK-level used in the summation.
%\begin{table}[H]
%\begin{center}
%\resizebox{16cm}{!}{
%\begin{tabular}{|c||c|c||c|c||c|c|}
%\hline 
%{}&\multicolumn{2}{|c||}{$R_V=1.0$} & \multicolumn{2}{c||}{$R_V=2.0$} &
%  \multicolumn{2}{c|}{$R_V=3.0$} \\
%\hline{$R_f$}&
%5 KK-level &  20 KK-level & 
%5 KK-level &  20 KK-level & 
%5 KK-level &  20 KK-level \\
%\hline
%1&0.586& 0.603&0.588  &0.610&0.590  &0.617\\
%2&0.672& 0.686&0.682  &0.700&0.693  &0.717\\
%3&0.734& 0.745&0.752  &0.765&0.766  &0.786\\
%4&0.780& 0.783&0.800  &0.810&0.820  &0.834\\
%\hline
%\end{tabular}
%}
%\end{center}
%\caption[]{Lower limits on $R^{-1}$ (in TeV) derived from $B_s$ decay branching ratio to $\mu^+ \mu^-$ for different values of input parameters showing the insensitivity on the number 
%of  KK-levels in summation (see Eq.\;\ref{cn} and Eq.\;\ref{bn}).}
%\label{sum-kk}
%\end{table}

\hspace*{-2cm}
\begin{table}[H]
%\hspace*{-5cm}
\begin{center}
\resizebox{18cm}{!}{
\begin{tabular}{|c||c|c||c|c||c|c||c|c||c|c||c|c|}
\hline 
{}&\multicolumn{2}{|c||}{$R_V=-1$}&\multicolumn{2}{|c||}{$R_V=0$} &\multicolumn{2}{|c||}{$R_V=1$} & \multicolumn{2}{c||}{$R_V=2$} &
  \multicolumn{2}{c|}{$R_V=3$} &  \multicolumn{2}{c|}{$R_V=4$}\\
\hline{$R_f$}&
5 KK-level &  20 KK-level &
5 KK-level &  20 KK-level &
5 KK-level &  20 KK-level & 
5 KK-level &  20 KK-level &
5 KK-level &  20 KK-level & 
5 KK-level &  20 KK-level \\
\hline
-2&0.333&0.357&0.246&0.260&0.119& 0.122&0.077  &0.081&0.045  &0.050&0.035&0.039\\
-1&0.361&0.378&0.280&0.300&0.191& 0.192&0.146  &0.150&0.119  &0.120&0.101&0.105\\
 0&0.368&0.380&0.454&0.460&0.412& 0.418&0.392  &0.402&0.383  &0.392&0.373&0.384\\
 1&0.381&0.399&0.553&0.567&0.586& 0.603&0.588  &0.610&0.590  &0.617&0.593&0.621\\
 2&0.399&0.406&0.617&0.624&0.672& 0.686&0.682  &0.700&0.693  &0.717&0.698&0.723\\
 3&0.403&0.409&0.661&0.664&0.734& 0.745&0.752  &0.765&0.766  &0.786&0.770&0.792\\
 4&0.410&0.416&0.694&0.701&0.780& 0.783&0.800  &0.810&0.820  &0.834&0.825&0.838\\
\hline
\end{tabular}
}
\end{center}
\caption[]{Lower limits on $R^{-1}$ (in TeV) derived from $B_s$ decay branching ratio to $\mu^+ \mu^-$ for different values of input parameters showing the insensitivity on the number 
of  KK-levels in summation (see Eq.\;\ref{cn} and Eq.\;\ref{bn}).}
\label{sum-kkn}
\end{table}

\subsection{Electroweak precision constraints}

Before we conclude, it would be relevant to discuss the constraints on the parameters coming from  our analysis vis-a-vis electroweak data, which remains very instrumental in constraining any new-physics beyond the SM.  Electroweak precision constraints on nmUED parameters have been discussed 
previously in \cite{flacke_kong, flacke_Pasold}, while authors in ref.\;\cite{flacke} restricts nmUED parameters from the consideration of $Z$-mass. However, in both of these cases nmUED action used for analysis are slightly different from that of ours. Ref.\;\cite{flacke_kong}, while presenting their electroweak results used {\em equal} BLKT parameters for all the field along with a bulk mass term for the fermions.  While the authors in the ref.\;\cite{flacke},
have considered several choices for the BLKT parameters. The case closest to our approach is where they have used equal BLKT parameters for Higgs and gauge bosons and set the BLKT parameter for the fermions equals to zero. 
Consequently, it would not be meaningful to directly apply the constraints derived in the above analysis to our case.   So we have redone the analysis following the approaches of refs.\;\cite{flacke, flacke_kong}, however applied to our 
case.

In nmUED model,  corrections to Peskin-Takeuchi parameters $S$, $T$ and $U$  as well as to $Z$ mass appear through the correction to the Fermi constant, $G_F$ at tree level, which is in stark contrast to 
the minimal version of the UED model where 
these correction appears through one loop processes. These quantities are modified by the correction of Fermi constant $G_F$ which is determined from muon
decay, i.e. a four-fermion process. The corrected Fermi constant $G_F$ can be decomposed into two parts as:
\begin{equation}
G_F=G_F^0+\delta G_F,
\end{equation}
with $G_F^0$ is simply the contribution from $W^\pm$ 0-mode exchange, while $\delta G_F$ denotes the sum of
the contributions from all non zero (even) $W^\pm$ KK-modes. The effective
Fermi constant can be expressed as the following.
\begin{equation}
G_F^0=\frac{g^2_2}{4\sqrt{2}M^2_W},~~~~~\delta G_F=\sum_{n\geq2}\frac{g^2_2(I^G_{n00})^2}{4\sqrt{2}M^2_{W^{(n)}}},
\end{equation}
where,
\begin{eqnarray}
I^G_{n00}&=&\sqrt{\pi R\left(1+\frac{r_V}{\pi R}\right)}\int_0 ^{\pi R}
dy \; \left[1 + r_{f}\{ \delta(y) + \delta(y - \pi R)\}\right]a^nf_L^0f_L^0,\\ \nonumber
&=&\frac{\sqrt{(1+\frac{r_V}{\pi R})}}{(1+\frac{r_f}{\pi R})}\frac{\sqrt{2}}{{\sqrt{1 + \frac{r^2_V m^2_{V^{(n)}}}{4} + \frac{r_V}{\pi R}}}}\frac{(r_f-r_V)}{\pi R}.
\end{eqnarray}

A similar expression for $\delta G_F$ has been presented in ref.\;\cite{flacke}. The above expression for $\delta G_F$, agrees with the same in ref.\;\cite{flacke} in the limit $r_f \rightarrow 0$.

Note that the above integral becomes zero when $r_f=r_V$, i.e., there is no correction of Fermi constant for this equality condition. 

 nmUED contributions to the $S$, $T$ and $U$ parameters can be written, following the approach of ref.\;\cite{flacke_kong, flacke_Pasold}, as:

\begin{equation}
S_{\rm nmUED}=0,~~T_{\rm nmUED}=-\frac{1}{\alpha}\frac{\delta G_F}{G_F},~~U_{\rm nmUED}=\frac{4 \sin^2 \theta_{w}}{\alpha}\frac{\delta G_F}{G_F}.
\end{equation}

One can now compare the predictions from nmUED model with the experimental values given in the ref.\;\cite{gfit}
\begin{equation}
S_{\rm NP} = 0.05 \pm 0.11,~~~T_{\rm NP} = 0.09 \pm 0.13,~~~ U_{\rm NP} = 0.01 \pm 0.11,
\label{stu_value}
\end{equation}
with input Higg mass $m_h = 125$ GeV and top quark mass $m_t = 173$ GeV. This would constrain the parameter space of nmUED model.

Before presenting the results of the above analysis, we would briefly discuss how $\delta G_F$ modifies the  $Z$-boson mass at the tree level \cite{flacke}.
 This would be evident once we express the  mass of $Z$-boson in terms  of $\alpha$, $M_W$ and $G_F$ : $M_Z=\frac{M_W\sqrt{\sqrt{2}G_F M^2_W}}{\sqrt{\sqrt{2}M^2_W G_F-\pi \alpha}}$.
 The corrections to $Z$ mass at tree level, in the framework of nmUED, would creep in through the corrected Fermi-constant, $G_F$. One can compare this with the so called tree level $Z$-mass defined by 
 the relation $m_Z ^{(expt)} - \delta ^{(1-loop)}m_Z$ following the ref.\;\cite{flacke}. Here, $\delta ^{(1-loop)}m_Z$ stands for the 1-loop correction to the $Z$-boson mass in the SM.

%\begin{figure}[t]
%\vspace*{-3cm}
%\begin{center}
%\includegraphics[width=5.5cm,height=5.5cm, angle =0]{Br_s_rg1_big}
%\includegraphics[width=5.5cm,height=5.5cm, angle =0]{Br_s_rg2_big}
%\includegraphics[width=5.5cm,height=5.5cm, angle =0]{Br_s_rg3_big}
%\includegraphics[scale=1.05,angle=0]{Precision_rv-1}
%\includegraphics[scale=1.05,angle=0]{Precision_rv1}
%\includegraphics[scale=1.05,angle=0]{Precision_rv2}
%\includegraphics[scale=1.05,angle=0]{Precision_rv3}
%\caption{Variation of $M_Z$ with $R^{-1}$ for several values of $R_f=r_f/R$ . The three panels correspond to different $R_V=r_V/R$.}
%\label{Precision}
%\end{center}
%\end{figure}

\begin{figure}[H]
\begin{center}
\includegraphics[scale=1.1,angle=0]{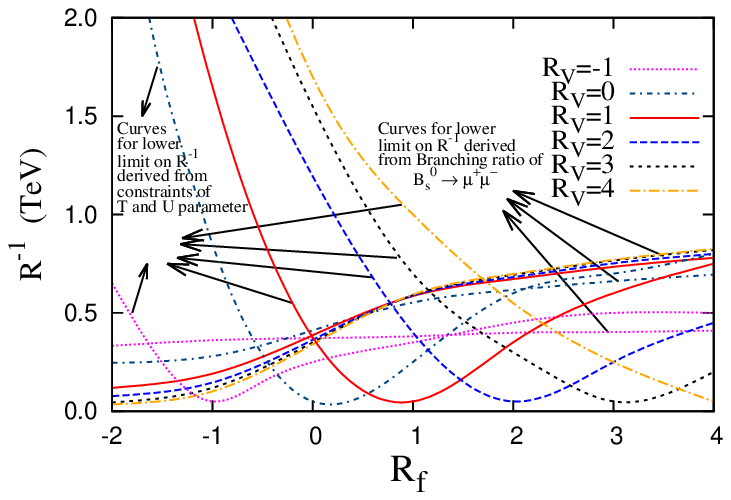}
\includegraphics[scale=1.1,angle=0]{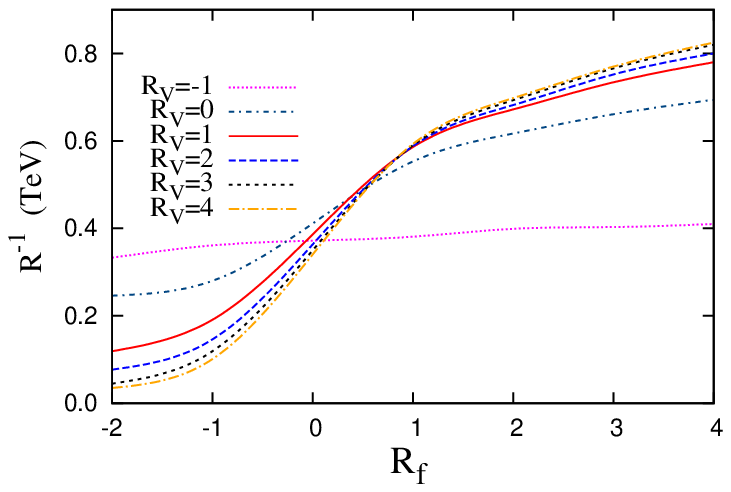}
\caption{95\% C.L. exclusion contours in $R_f - R^{-1}$ plane for six different choices of $R_V$ from branching ratio of $B_s \rightarrow \mu^+ \mu^-$ decay and from constraints of $T$ and $U$ parameter. The area below a particular curve (fixed $R_V$) has been excluded at 95\% C.L.}
\label{cont}
\end{center}
\end{figure}

We present our results in Fig.\;\ref{cont} (left panel) in terms of 95\% C.L. lower limit on $R^{-1}$ for several values of BLKT parameters. Region of the parameter space below  a particular line has been ruled out from the consideration of $T$ and $U$ parameters. Furthermore, we have checked that the {\em tree level value} of $Z$-boson mass in the framework of nmUED, resides within the 95 \% C.L.  allowed tree level value \cite{flacke} for the entire range of parameter space that we have considered in this work. 

%begin{tabular}{|c|c|c|c|c|c|}
%\hline 
%{$R_f$}&{$R_V=-1.0$}&{$R_V=1.0$}&{$R_V=2.0$}&{$R_V=3.0$}&{$R_V=4.0$}\\
%\hline
%-2&0.65&-&-&-&-\\ \hline
%-1&0.05&1.65&2.20&2.70&-\\ \hline
 %1&0.35&0.05&0.40& 0.70&1.00\\ \hline
 %2&0.45&0.35&0.05& 0.30&0.55\\ \hline
 %3&0.50&0.60&0.25& 0.05&0.25\\ \hline
 %4&0.50&0.75&0.45& 0.20&0.05\\ \hline
%\end{tabular}
%\end{center}
%\caption{Lower limits on $R^{-1}$ (in TeV) derived from $T$ and $U$ parameter for different values of input parameters.}
%\label{precision_box}
%\end{table}

To compare the limits derived from our analysis ($B_s \rightarrow \mu^+ \mu^-$) with that from electroweak precision test,
we present the allowed range of parameter space both from $B_s$ decay and electroweak precision data,  in $R_f - R^{-1}$ plane for six different choices of $R_V$  in Fig.\;\ref{cont} (left panel). In the right panel of Fig.\;\ref{cont}, we present the allowed parameter space, on an expanded scale,
only from the $B_s$ decay  analysis for sake of clarity. Each curve (in the left and right panel) corresponds to a particular value of $R_V$, each point of which represents a lower limit of $R^{-1}$ in TeV corresponding to a particular value of $R_f$. So that the area under a particular curve has been dis-allowed by either from precision constraints or from $Br(B_s \rightarrow \mu^+ \mu^-)$. It is evident from the plots that 
for a given value of $R_V$, higher values of $R^{-1}$ are being ruled out from electroweak data for lower values of $R_f$, while $B_s$
branching ratio would do better than the electroweak data in excluding values of $R^{-1}$ for higher values of $R_f$.  Finally, we mention while passing that the all of the parameter space (whether allowed or disallowed from electroweak data and $B_s$ decay) is consistent with the $Z$-boson mass following the ref.\;\cite{flacke}.

It has been evident from our analysis that the lower bounds on $R^{-1}$ derived with negative values of BLKT parameters are not very interesting and the above values of $R^{-1}$ have been already excluded from the consideration of electroweak data, which can be 
seen from the left panel of Fig.\;\ref{cont}.  The weak nature of the bounds on $R^{-1}$ derived from $B_s \rightarrow \mu^+ \mu^-$ 
branching ratio for negative values of BLT parameters, could be accounted by the higher masses of the KK-excitations, diminishing the contributions of penguins and boxes to the total decay amplitude. Furthermore, we have restricted  our choice of BLT parameters upto
4. A careful look at the left or right panel  of Fig.\;\ref{cont} (or Table\;\ref{sum-kkn}), tells us that lower limits on $R^{-1}$ are weakly sensitive to the 
values of $R_V$. So we expect that higher values of  $R_V$ would keep the lower limit on $R^{-1}$ in the same ballpark.

\section{Conclusion}

We have calculated the contribution of KK-excitations in the framework of non-minimal Universal Extra Dimensional model  to the branching ratio of $B_{s(d)} \rightarrow \mu^+ \mu^-$ at one loop level. Non-minimal UED is hallmarked by the presence 
of boundary localised kinetic and  Yukawa terms along with an SM like action and field content however defined  in a $4 + 1$ dimensional bulk. Boundary 
localised terms parametrise the unknown radiative corrections
to the masses and couplings in the full 5-D theory. Presence of boundary terms modify the couplings and mass spectrum of KK-modes in the 4-D effective theory in a non-trivial manner.

To put our discussion into a context, we must remind that in UED, masses of the $n^{th}$ KK-modes are $n R^{-1}$  ($R^{-1}$ being the compactification scale). Interactions among the different KK-excitations are 
very similar to their SM (0-modes) counterparts. However inclusion of BLT parameters would shift the masses of KK-modes from their UED values. For keeping our analysis simple we stick to the case of 
two different BLT parameters. The first one $r_V$,  specifies the gauge and Higgs BLTs while the second one, $r_f$ stands for equal fermion and Yukawa BLT coefficients. 

Effective interaction for $B_{s(d)}$ meson decaying to a pair of $\mu^+ \mu^-$ can be parametrised by  a 4-fermion interaction. The coefficient for this effective interaction is calculable in the framework of SM and in the model of our 
interest namely the UED model. There are two sets of diagrams contributing to this decay process. The $Z$-penguins contribute dominantly while the box diagrams are sub-dominant.  Diagrams 3, 4, 6, 7 and 8 in Fig.\;\ref{pen} and Ib and IIb of Fig.\;\ref{self} captures the effects of top-Yukawa couplings, due to which this one loop mediated process is being amplified.

We have listed the amplitudes from each diagram separately in Appendix A and B.  The total contribution coming from the penguins are finite and GIM mechanism plays a crucial role to tame the divergences. It should be noted that the final results is  sum of contributions coming from different KK-levels and it also contains the SM contribution (i.e. from $0^{th}$ KK-mode). While 
summing over the KK-levels we restrict ourselves to 5 levels in view of a recent analysis relating the Higgs boson mass and cut-off of a UED theory \cite{kksum}. It has been shown explicitly that the limits on the parameters derived from our analysis would change a little if we take 20 KK-levels in the summation instead of 5.

There is a one (two)-standard deviation difference between the experimentally measured branching ratio of 
$B_{s(d)} \rightarrow \mu^+ \mu^-$ with its SM prediction. We have used the experimental data  on the measured value of branching 
ratio of $B_{s(d)} \rightarrow \mu^+ \mu^-$ to  constrain the parameter space of nmUED model. As our calculation would 
reproduce the results of $Br(B_s \rightarrow \mu^+ \mu^-)$ in the vanishing BLKT limits, at the very outset, we have used the 
above algorithm to set a lower limit on $R^{-1}$ in the framework of UED. 
In case of UED, present analysis constrained $R^{-1}$ to greater than 450 GeV at 95 \% C.L. This limit in the framework of UED, 
is not one of the most stringent.  However, this is in the same ballpark  with the limits those are obtained from the consideration of $R_b$ \cite{zbb} or $\rho$-parameters \cite{rho}. The nmUED model has more than one parameters apart from the compactification radius. 
The BLT coefficients control the masses of KK-modes and their interactions. So in the nmUED framework, any lower limit on $R^{-1}$ 
would depend on these BLT coefficients. As for example, for $R_f = R_V = 3$, $R^{-1} >0.7 \rm\;TeV$, while for $R_f = 4, 
\; R_V = 3$, $R^{-1} > 0.8 \rm\;TeV$. These lower limits on  $R^{-1}$ are so far the most stringent  one in the framework of 
nmUED. Thus the recent experimental result of $B_s$ meson decay to $\mu^+ \mu^-$, combined with present analysis has 
excluded the largest region in the nmUED parameter space. Unfortunately the bounds on the parameters using the 
$Br(B_d \rightarrow \mu^+ \mu^-)$ is not so competitive.  Lower limits on $R^{-1}$ derived from our analysis for negative 
values of $R_f$ are not so stringent and the have been already ruled out from the consideration of electroweak precision
data.    

We have also reviewed the effect of electroweak precision constraint on nmUED parameter space.  Our analysis reveals that the 
experimental data on the $B_s \rightarrow \mu^+ \mu^-$ branching ratio would put more severe constraint on the lower limit of 
$R^{-1}$ for positive values of $R_f$ than the electroweak precision data.\\

{\bf Acknowledgements} AD is partially supported by DAE-BRNS research project. AS acknowledges financial support from UGC 
in form of a Senior Research Fellowship. Authors are grateful to Sreerup Raychaudhury for taking part in the initial stage and many useful discussion.

\newpage
\begin{appendices}
\renewcommand{\thesection}{\Alph{section}}
\renewcommand{\theequation}{\thesection-\arabic{equation}} 

\setcounter{equation}{0}  

\section{Different contributions to \boldmath{$Z$}-penguin and self-energy diagrams}
The contributions of the $Z$-penguin diagrams to the functions $C_n$ in Fig.\;\ref{pen} are given as follows:
\begin{eqnarray}
F_{1}(x_{f(n)}) \!\!\!\!&=&\!\!\!\!\frac{1}{8}\left(c^4_{fn}+s^4_{fn}-\frac{4}{3} s^2_w\right)
%\nonumber \\ &&
\left[ \Delta +\ln\frac{\mu^2}{M^2_{f^{(n)}}}-\frac{3}{2}+h_q\left(x_{f(n)}\right)-2 x_{f(n)} h_q\left(x_{f(n)}\right)\right](I^n_1)^2,\\
F_{2}(x_{f(n)}) \!\!\!\!&=& \!\!\!\!\frac{1}{4}c^2_{fn}s^2_{fn}\left[ \Delta +\ln\frac{\mu^2}{M^2_{f^{(n)}}}-\frac{3}{2}+h_q\left(x_{f(n)}\right)+2 x_{f(n)} h_q\left(x_{f(n)}\right)\right](I^n_1)^2,\\
F_{3}(x_{f(n)}) \!\!\!\!&=&\!\!\!\!\frac{1}{16 M^2_{W^{(n)}}}
  \Bigg[ \left( (m^{{(f)}}_{1})^2+ (m^{{(f)}}_{3})^2\right)\left(c_{f{n}}^{2} -\frac{4}{3} s^2_w \right)
%\nonumber \\ &&
    +\left((m^{{(f)}}_{2})^2+ (m^{{(f)}}_{4})^2\right)  \left(s_{fn}^2-\frac{4}{3} s^2_w \right)\Bigg]
\nonumber \\ &&
\times\left[ \Delta + 
      \ln \frac{\mu^2}{M^2_{f^{(n)}}}-\frac{1}{2}+h_q\left(x_{f(n)}\right) 
      -2 x_{f(n)} h_q\left(x_{f(n)}\right)\right],\\
F_{4}(x_{f(n)})\!\!\!\!&=& \!\!\!\!-\frac{1}{8 M^2_{W^{(n)}}} 
      \left( m^{{(f)}}_{1}\, m^{{(f)}}_{2}
    + m^{{(f)}}_{3}\, m^{{(f)}}_{4}\right)~c_{fn}~s_{fn} 
   \nonumber \\ &&
\times\left[ \Delta + 
  \ln\frac{\mu^2}{M^2_{f^{(n)}}}-\frac{1}{2}  +  h_q\left(x_{f(n)}\right) 
    + 2 x_{f(n)} h_q\left(x_{f(n)}\right)\right],\\
F_{5}(x_{f(n)})\!\!\!\!&=&\!\!\!\! -\frac{3}{4}c^2_w\left[\Delta + \ln \frac{\mu^2}{M^2_{W^{(n)}}}-\frac{1}{6}-x_{f(n)}h_w\left(x_{f(n)}\right)\right](I^n_1)^2,\\
%\end{eqnarray}
%\begin{eqnarray}
F_{6}(x_{f(n)})\!\!\!\!&= &\!\!\!\!-\frac{1}{16\, M^{4}_{W^{(n)}}} 
  \Bigg[ \bigg( \left( 1-2\,s^2_w \right) M_W^2 + 2 c^2_w m^2_{V^{(n)}} \bigg) 
     \left((m^{{(f)}}_{1})^2+ (m^{{(f)}}_{2})^2\right)  
 \nonumber \\ && + \bigg(\left(1-2\,s^2_w \right) m^2_{V^{(n)}}   + 2 c^2_w M_W^2 \bigg)  \left((m^{{(f)}}_{3})^2+ (m^{{(f)}}_{4})^2\right)
        \Bigg]
 \nonumber \\ &&\times\left[ \Delta + 
  \ln \frac{\mu^2}{M_{W(n)}^2}+\frac{1}{2}  - x_{f(n)} h_w\left(x_{f(n)}\right) \right],
\\
F_{7}(x_{f(n)}) \!\!\!\!&=&\!\!\!\!\frac{M_W  m_{V^{(n)}}}{8 M^{4}_{W^{(n)}}}
 \left( m^{{(f)}}_{1} m^{{(f)}}_{3} +  m^{{(f)}}_{2}
  m^{{(f)}}_{4}\right) 
%\nonumber \\&& 
  \left[ \Delta + 
    \ln \frac{\mu^2}{M_{W(n)}^2}+\frac{1}{2}  - x_{f(n)}\, h_w\left(x_{f(n)}\right)\right],\\
%\end{eqnarray}
%\begin{eqnarray}
F_{8}(x_{f(n)})\!\!\!&=&\!\!\!\frac{M_{f^{(n)}} }{2 M^{4}_{W^{(n)}}}
  \Bigg[ \bigg(s^2_w M_W^2 - c_w^2 m^2_{V^{(n)}}\bigg)\left(m^{{(f)}}_{1} ~c_{fn} +m^{{(f)}}_{2} ~s_{fn} \right)
\nonumber \\&& 
+ M_W
      m_{V^{(n)}}\left( m^{{(f)}}_{3} ~c_{fn} 
       + m^{{(f)}}_{4} ~s_{fn} \right)\Bigg]\,
       h_w\left(x_{f(n)}\right)I^n_1,
\end{eqnarray}
where the functions $h_q$ and $h_w$ are given by:
\begin{eqnarray}
    h_q(x) &=& \frac{1}{1-x} + \frac{\ln x}{(1-x)^2}, \\
    h_w(x) &=& \frac{1}{1-x} + \frac{x \ln x}{(1-x)^2} ~.
  \end{eqnarray}
The contributions of the self-energy diagrams to the functions $C_n$ are given by the following expressions which are generated from Fig.\;\ref{self}:
{\small
 \begin{eqnarray}
     \Delta S_1\left(x_{f(n)} \right) \!\!\! &=&  \!\!\!\frac14 \left[\Delta
    -\frac{1}{2} \left\{\frac{1+x_{f(n)}}{1-x_{f(n)}} +  \frac{2
    x_{f(n)}^2 \ln x_{f(n)} }{(1-x_{f(n)})^2} \right\} - 
    \ln { \frac{M^2_{W^{(n)}}}{\mu^2}} \right](I^n_1)^2,\\
     \Delta S_2\left(x_{f(n)} \right)\! \!\! &=& \! \!\!\frac18 \left[
    (I^n_2)^2+ (I^n_1)^2x_f \right]  \left[\Delta +\frac{1}{2} 
\left\{ \frac{1-3x_{f(n)}}{1-x_{f(n)}} -
          \frac{2 x_{f(n)}^2\ln x_{f(n)} }{(1-x_{f(n)})^2} \right\}   - 
    \ln { \frac{M^2_{W^{(n)}}}{\mu^2}} \right] \!.
  \end{eqnarray}
}
%  \begin{figure}[hbt] \centering
%  \subfigure[]{
%        \includegraphics[]{self1.eps} }\hspace{0.4cm}
%   \subfigure[]{
%        \includegraphics[]{self2.eps} }
 
%    \caption[]{\small\label{selfenergydiagrams} Self-energy diagrams
%    necessary for calculating the electroweak counter term as
%    discussed in \cite{BB1}.}
%  \end{figure}
Here $\Delta=\frac{2}{\epsilon}+\ln 4\pi-\gamma_{E}$, \;\; $D=4-\epsilon$.
\setcounter{equation}{0} 
\section{Different contributions to box diagrams}
The contributions of the box diagrams to the functions $B_n$ in Fig.\;\ref{box} are given as follows:
\begin{eqnarray}
    H_{WW(n)} &=&-\frac14 \frac{M_W^2}{M^2_{W^{(n)}}}~ U(x_{f(n)}, x_{\nu(n)})(I^n_1)^4,\\
    H_{WG(n)} &=& \frac12\,\frac{M_W^2 M_{f^{(n)}}m_{\nu^{(n)}}}{{M^6_{W^{(n)}}}} \left[
      m_1^{(f)} c_{f(n)} + m_2^{(f)} s_{f(n)} \right] ~m_1^{(\nu)}~
      \widetilde U(x_{f(n)}, x_{\nu(n)})(I^n_1)^2~, \\
       H_{WH(n)} &=& \frac12\,\frac{M_W^2 M_{f^{(n)}}m_{\nu^{(n)}}}{{M^6_{W^{(n)}}}} \left[
      m_3^{(f)} c_{f(n)} + m_4^{(f)} s_{f(n)} \right]~ m_3^{(\nu)} ~
      \widetilde U(x_{f(n)}, x_{\nu(n)})(I^n_1)^2~, \\
      H_{GH(n)} &=& -\frac{1}{8} \frac{M_W^2}{{M^6_{W^{(n)}}}}
    \left[m_1^{(f)} m_3^{(f)} + m_2^{(f)}
      m_4^{(f)}\right]~m_1^{(\nu)}\, m_3^{(\nu)}~ U(x_{f(n)}, x_{\nu(n)})~,\\
     H_{GG(n)} &=&-\frac{1}{16} \frac{M_W^2}{{M^6_{W^{(n)}}}}
   \left[(m_1^{(f)})^2 + (m_2^{(f)})^2 \right] (m_1^{(\nu)})^2~  U(x_{f(n)}, x_{\nu(n)})~,\\
    H_{HH(n)} &=& -\frac{1}{16}\frac{M_W^2}{{M^6_{W^{(n)}}}}
   \left[(m_3^{(f)})^2 + (m_4^{(f)})^2\right] (m_3^{(\nu)})^2 ~  
U(x_{f(n)}, x_{\nu(n)})~.
 \end{eqnarray}

The functions $U$ and $\widetilde U$ are defined as,

\begin{eqnarray}\label{U1}
    U (x_t, x_u) &=&\frac{x_t^2\log{x_t}}{(x_t-x_u)(1-x_t)^2} +
    \frac{x_u^2\log{x_u}}{(x_u-x_t)(1-x_u)^2} + \frac{1}{(1-x_u)(1-x_t)}~,\\ 
  \label{U3}
    \widetilde U (x_t, x_u) &=& \frac{x_t\log{x_t}}{(x_t-x_u)(1-x_t)^2} +
    \frac{x_u\log{x_u}}{(x_u-x_t)(1-x_u)^2} + \frac{1}{(1-x_u)(1-x_t)}~.
\end{eqnarray}

%\vspace{.5cm}
%\begin{minipage}{38mm}
%  \fbox{\includegraphics[]{vertex1.eps}}
%\end{minipage}
%\begin{minipage}{12cm}
%\newpage
\setcounter{equation}{0} 
\section{Feynman rules for \boldmath{$B_{s(d)}\rightarrow\mu^{+}\mu^{-}$} in nmUED}

The Feynman rules for the different vertices with the assumption that all momenta and fields are incoming. 

1) $Z^{\mu}W^{\nu\pm}S^{\mp}$
$\displaystyle : \frac{g_2}{c_w M_{W^{(n)}}} g_{\mu\nu} C$, where $C$ is given by:
%\end{minipage}

\begin{equation}
 \begin{aligned}
  Z^{\mu} W^{(n)+} G^{(n)-}: C   &= - s^2_w M_W^2 + c^2_w m^2_{V^{(n)}},\\ 
  Z^{\mu} W^{(n)-} G^{(n)+}: C   &= s^2_w M_W^2 - c^2_w m^2_{V^{(n)}},\\
  Z^{\mu} W^{(n)+} H^{(n)-}: C   &= -iM_W m_{V^{(n)}},\\
  Z^{\mu} W^{(n)-} H^{(n)+}: C   &= iM_W m_{V^{(n)}}.  
 \end{aligned}
\end{equation}

%\vspace{.2cm} 
%\begin{minipage}{42mm}
%  \fbox{\includegraphics[]{vertex2.eps}}
%\end{minipage}
%\begin{minipage}{12cm}
2) $Z^{\mu}S^{\pm}_1S^{\mp}_2$
$\displaystyle : \frac{ig_2}{2c_w M^2_{W^{(n)}}} (k_2-k_1)_{\mu} C$, where $C$ is given by:
%\end{minipage}

\begin{equation}
 \begin{aligned}
  Z^{\mu} G^{(n)+} G^{(n)-}: C &= -\left(c^2_w-s^2_w\right) M_W^2 - 2c^2_w m^2_{V^{(n)}},\\   
  Z^{\mu} H^{(n)+} H^{(n)-}: C &=  - 2c^2_w M_W^2 -  \left(c^2_w-s^2_w\right)m^2_{V^{(n)}},\\
  Z^{\mu} G^{(n)+} H^{(n)-}: C &= iM_W m_{V^{(n)}},\\ 
  Z^{\mu} G^{(n)-} H^{(n)+}: C &= -iM_W m_{V^{(n)}}.
\end{aligned}
\end{equation}

Here the scalar fields $S\equiv H,G.$

%\vspace{.2cm}
%\begin{minipage}{44mm}
%  \fbox{\includegraphics[]{vertex3.eps}}
%\end{minipage}
%\begin{minipage}{12cm}
%\begin{equation}
3) $Z^{\mu}(k_1)W^{\nu+}(k_2)W^{\lambda-}(k_3)$
$\displaystyle :$
\begin{equation}
 ig_2c_w \left[ g_{\mu\nu} (k_2
      -k_1)_\lambda + g_{\mu\lambda} (k_1 -k_3)_\nu +
    g_{\lambda\nu} (k_3 -k_2)_\mu \right].
%\end{minipage}
\end{equation}

%\vspace{.2cm}
%\begin{minipage}{36mm}
 % \fbox{\includegraphics[]{vertex4.eps}}
%\end{minipage}
%\begin{minipage}{12cm}
4) $Z^{\mu}{\overline{f}_1} f_2$
  $\displaystyle  : \frac{i g_2}{6 c_w} \gamma_\mu \left( P_L C_L +
    P_R C_R \right)$, where $C_L$ and $C_R$ are given by:
%\end{minipage}

%\begin{alignat}{4}
\begin{equation}
\begin{aligned}
  & Z^{\mu} \bar{u_i} u_i:  &
  &\left\{\begin{array}{l}C_L = 3-4s^2_w,\\
      C_R = -4s^2_w,\end{array}\right.
  && Z^{\mu} \bar{d_j} d_j:    & 
  &\left\{\begin{array}{l}C_L = -3+2s^2_w,\\
      C_R = 2s^2_w,\end{array}\right.\\
  &Z^{\mu} \bar {\nu_i} \nu_i:  & 
  &\left\{\begin{array}{l}C_L = 3,\\
      C_R = 0,\end{array}\right.
  && Z^{\mu} \bar {e_j} e_j:     &
  &\left\{\begin{array}{l}C_L = -3+6s^2_w,\\
      C_R = 6s^2_w ,\end{array}\right.\\
  & Z^{\mu} {\overline{T}^{1(n)}_i} T^{1(n)}_i: &   
  &\left\{\begin{array}{l}C_L = -4s^2_w+3c^2_{in},\\
      C_R = -4s^2_w+3c^2_{in},\end{array}\right.
  && Z^{\mu} {\overline{T}^{2(n)}_i} T^{2(n)}_i:     &
  &\left\{\begin{array}{l}C_L = -4s^2_w+3s^2_{in},\\
      C_R = -4s^2_w+3s^2_{in},\end{array}\right.\\
  & Z^{\mu} {\overline{T}^{1(n)}_i} T^{2(n)}_i:  &    
  &\left\{\begin{array}{l}C_L = -3s_{in} c_{in},\\
      C_R = 3s_{in} c_{in},\end{array}\right.
  &&Z^{\mu} {\overline{T}^{2(n)}_i} T^{1(n)}_i:     & 
  &\left\{\begin{array}{l}C_L = -3s_{in} c_{in},\\
      C_R = 3s_{in} c_{in}.\end{array}\right.
\end{aligned}
\end{equation}
%
%\end{alignat}

5) $S^{\pm}{\overline{f}_1} f_2$
  $\displaystyle  = \frac{g_2}{\sqrt{2} M_{W^{(n)}}} (P_L C_L + P_R C_R)$, where $C_L$ and $C_R$ are given by:
%\end{minipage}

%\begin{alignat}{4}
\begin{equation}
\begin{aligned}
  & G^+ \bar{u_i} d_j :  &
  &\left\{\begin{array}{l}C_L = -m_i V_{ij},\\
      C_R = m_j V_{ij},\end{array}\right.
  &&G^- \bar{d_j} u_i :     &
  &\left\{\begin{array}{l}C_L = -m_j V_{ij}^*,\\
      C_R = m_i V_{ij}^*,\end{array}\right.\\
  & G^{(n)+}{\overline{T}^{1(n)}_i} d_j :  &
  &\left\{\begin{array}{l}C_L = -m_1^{(i)} V_{ij},\\
      C_R = M_1^{(i,j)} V_{ij},\end{array}\right.
  &&G^{(n)-}\bar{d_j}T^{1(n)}_i :   &
  &\left\{\begin{array}{l}C_L = -M_1^{(i,j)} V_{ij}^*,\\
     C_R = m_1^{(i)} V_{ij}^*,\end{array}\right.\\
  & G^{(n)+}{\overline{T}^{2(n)}_i} d_j :  &
  &\left\{\begin{array}{l}C_L = m_2^{(i)} V_{ij},\\
      C_R =-M_2^{(i,j)} V_{ij},\end{array}\right.
  &&G^{(n)-}\bar{d_j}T^{2(n)}_i :   &
  &\left\{\begin{array}{l}C_L = M_2^{(i,j)} V_{ij}^*,\\
     C_R =-m_2^{(i)} V_{ij}^*,\end{array}\right.\\
  & G^+ \bar{\nu_i} e_j :  &
  &\left\{\begin{array}{l}C_L = 0,\\
      C_R = m_j \delta_{ij},\end{array}\right.
  &&G^- \bar{e_j} \nu_i :     &
  &\left\{\begin{array}{l}C_L = -m_j \delta_{ij},\\
      C_R = 0,\end{array}\right.\\
  & G^{(n)+} \bar{\nu_i} {\mathcal L}^{(n)}_j :  &
  &\left\{\begin{array}{l}C_L = 0,\\
      C_R = m^{(j)}_1 \delta_{ij},\end{array}\right.
  &&G^{(n)-}{\overline{\mathcal L}^{(n)}_j} \nu_i :     &
  &\left\{\begin{array}{l}C_L = -m^{(j)}_1 \delta_{ij},\\
      C_R = 0,\end{array}\right.\\
  & G^{(n)+} \bar{\nu_i} {\mathcal E}^{(n)}_j :  &
  &\left\{\begin{array}{l}C_L = 0,\\
      C_R =-m^{(j)}_2 \delta_{ij},\end{array}\right.
  &&G^{(n)-}{\overline{\mathcal E}^{(n)}_j} \nu_i :     &
  &\left\{\begin{array}{l}C_L = m^{(j)}_2 \delta_{ij},\\
      C_R = 0,\end{array}\right.\\
  & H^{(n)+}{\overline{T}^{1(n)}_i} d_j :  &
  &\left\{\begin{array}{l}C_L = -m_3^{(i)} V_{ij},\\
      C_R = M_3^{(i,j)} V_{ij},\end{array}\right.
  &&H^{(n)-}\bar{d_j}T^{1(n)}_i :   &
  &\left\{\begin{array}{l}C_L = -M_3^{(i,j)} V_{ij}^*,\\
     C_R = m_3^{(i)} V_{ij}^*,\end{array}\right.\\
  & H^{(n)+}{\overline{T}^{2(n)}_i} d_j :  &
  &\left\{\begin{array}{l}C_L = m_4^{(i)} V_{ij},\\
      C_R =-M_4^{(i,j)} V_{ij},\end{array}\right.
  &&H^{(n)-}\bar{d_j}T^{2(n)}_i :   &
  &\left\{\begin{array}{l}C_L = M_4^{(i,j)} V_{ij}^*,\\
     C_R =-m_4^{(i)} V_{ij}^*,\end{array}\right.\\
  & H^{(n)+} \bar{\nu_i} {\mathcal L}^{(n)}_j :  &
  &\left\{\begin{array}{l}C_L = 0,\\
      C_R = m^{(j)}_3 \delta_{ij},\end{array}\right.
  &&H^{(n)-}{\overline{\mathcal L}^{(n)}_j} \nu_i :     &
  &\left\{\begin{array}{l}C_L = -m^{(j)}_3 \delta_{ij},\\
      C_R = 0,\end{array}\right.\\
  & H^{(n)+} \bar{\nu_i} {\mathcal E}^{(n)}_j :  &
  &\left\{\begin{array}{l}C_L = 0,\\
      C_R =-m^{(j)}_4 \delta_{ij},\end{array}\right.
  &&H^{(n)-}{\overline{\mathcal E}^{(n)}_j} \nu_i :     &
  &\left\{\begin{array}{l}C_L = m^{(j)}_4 \delta_{ij},\\
      C_R = 0,\end{array}\right.\\
\end{aligned}
\end{equation}

Here the fermion fields $f\equiv u, d, T^1_t, T^2_t, \nu, e, {\mathcal L}_{\nu}, {\mathcal L}_e, {\mathcal E}.$
%\newpage

%\vspace{.5cm}
%\begin{minipage}{36mm}
%  \fbox{\includegraphics[]{vertex5.eps}}
%\end{minipage}
%\begin{minipage}{12cm}
6) $W^{\mu\pm}{\overline{f}_1}f_2$
  $\displaystyle  :  \frac{i g_2}{\sqrt{2}} \gamma_\mu P_L C_L$, where $C_L$ is given by:
%\end{minipage}
%\begin{alignat}{4}

\begin{equation}
\begin{aligned}
  & W^{\mu+}\bar{u_i} d_j : &&     C_L = V_{ij},
  && W^{\mu-}\bar{d_j} u_i : &&    C_L = V^*_{ij},\\
  & W^{\mu(n)+}{\overline{T}^{1(n)}_i}d_j : &&   C_L = I^n_1\;c_{in} V_{ij},
  &&W^{\mu(n)-}\bar{d_j}{{T}^{1(n)}_i} : && C_L = I^n_1\;c_{in} V^*_{ij},\\
  & W^{\mu(n)+}{\overline{T}^{2(n)}_i}d_j : &&   C_L = -I^n_1\;s_{in} V_{ij},
  &&W^{\mu(n)-}\bar{d_j}{{T}^{2(n)}_i} : && C_L = -I^n_1\;s_{in}V^*_{ij},\\
  & W^{\mu+}\bar {\nu_i} e_j : &&     C_L = \delta_{ij},
  && W^{\mu-}\bar{e_j} \nu_i : &&    C_L = \delta_{ij},\\
  & W^{\mu(n)+}\bar {\nu_i} {\mathcal L}^{(n)}_j : &&     C_L = I^n_1\;\delta_{ij},
  &\hspace{12ex} & W^{\mu(n)-}{\overline{\mathcal L}^{(n)}_j} \nu_i : &&    C_L =  I^n_1\;\delta_{ij},\\
  & W^{\mu(n)+}\bar {\nu_i} {\mathcal E}^{(n)}_j : &&     C_L = 0,
  && W^{\mu(n)-}{\overline{\mathcal E}^{(n)}_j} \nu_i : &&    C_L = 0.
\end{aligned}
\end{equation}

%\end{alignat}
%\newpage
%\vspace{.2cm}
%\begin{minipage}{36mm}
%  \fbox{\includegraphics[]{vertex6.eps}}
%\end{minipage}
%\begin{minipage}{12cm}

%\end{alignat}

The mass parameters $m_x^{(i)}$ are given by:
\begin{equation}
\label{mparameters}
  \begin{aligned}
    m_1^{(i)} &= I^n_2\;m_{V^{(n)}}c_{in} +I^n_1\;m_i s_{in},\\
    m_2^{(i)} &= -I^n_2\;m_{V^{(n)}}s_{in}+I^n_1\;m_i c_{in},\\
    m_3^{(i)} &= -I^n_2\;iM_W c_{in} +I^n_1\;i\frac{m_{V^{(n)}}m_i}{M_W}s_{in},\\
    m_4^{(i)} &= I^n_2\;iM_W s_{in}+I^n_1\;i\frac{m_{V^{(n)}}m_i}{M_W}c_{in},
  \end{aligned}
\end{equation}
where $m_i$ represents the mass of the 0-mode {\it up-type} fermion and $c_{in}=\cos(\alpha_{in})$ and $s_{in}=\sin(\alpha_{in})$ with $\alpha_{in}$ as defined earlier.

And the mass parameters $M_x^{(i,j)}$ are:
\begin{equation}\label{Mparameters}
  \begin{aligned}
    M_1^{(i,j)}  &= I^n_1\;m_j c_{in},\\
    M_2^{(i,j)}  &= I^n_1\;m_j s_{in},\\
    M_3^{(i,j)}  &= I^n_1\;i\frac{m_{V^{(n)}}m_j}{M_W}c_{in},\\
    M_4^{(i,j)}  &= I^n_1\;i\frac{m_{V^{(n)}}m_j}{M_W}s_{in},
  \end{aligned}
\end{equation}
where $m_j$ represents the mass of the 0-mode {\it down-type} fermion. Here, $I^n_1$ and $I^n_2$ are the overlap integrals given in Eqs.\;\ref{i1} and \ref{i2} respectively.

\end{appendices}


\begin{thebibliography}{99}
%\bibitem{ia} I. Antoniadis, Phys. Lett. B {\bf 246} (1990) 377.

\bibitem{ued_dm}G. Servant and T. M. P. Tait, New J. Phys. {\bf4} (2002) 99, [arXiv: hep-ph/0209262]; G.
Servant and T. M. P. Tait, Nucl. Phys. B {\bf650} (2003) 391, [arXiv: hep-ph/0206071]; H.
C. Cheng, J. L. Feng and K. T. Matchev, Phys. Rev. Lett. {\bf89} (2002) 211301, [arXiv:hep-
ph/0207125]; D. Majumdar, Phys. Rev. D {\bf67} (2003) 095010, [arXiv:hep-ph/0209277];
F. Burnell and G. D. Kribs, Phys. Rev. D {\bf73} (2006) 015001, [arXiv:hep-ph/0509118]; K.
Kong and K. T. Matchev, JHEP {\bf0601} (2006) 038 [arXiv:hep-ph/0509119]; M. Kakizaki,
S. Matsumoto and M. Senami, Phys. Rev. D {\bf74} (2006) 023504, [arXiv:hep-ph/0605280].

\bibitem {relic}G. Belanger, M. Kakizaki and A. Pukhov, JCAP {\bf 1102} (2011) 009, [arXiv:1012.2577 [hep-ph]].


\bibitem{ued_uni}K. R. Dienes, E. Dudas and T. Gherghetta, Phys. Lett. B {\bf436} (1998) 55, [arXiv:hep-ph/9803466]; K. Dienes, E. Dudas, and T. Gherghetta, Nucl. Phys. B {\bf537} (1999) 47, [arXiv:hep-ph/9806292]; G. Bhattacharyya, A. Datta, S. K. Majee and A. Raychaudhuri, Nucl. Phys. B {\bf760} (2007) 117, [arXiv:hep-ph/0608208].


\bibitem{hamed}K. Yoshioka, Mod. Phys. Lett. A {\bf15} (2000) 29, [arXiv:hep-ph/9904433]; P. R. Archer, JHEP {\bf 09} (2012) 095 [arXiv:1204.4730 [hep-ph]].
%B.~A.~Dobrescu and E. Poppiz, Phys. Rev. Lett. {\bf 87} (2001) 031801, [arXiv:hep-ph/0102010].

\bibitem{acd} T. Appelquist, H. C. Cheng and B. A. Dobrescu, 
 %``Bounds on universal extra dimensions,'' 
 Phys.\ Rev.\ D {\bf 64} (2001) 035002, [arXiv:hep-ph/0012100].
\bibitem{rad_cor_georgi}H. Georgi, A. K. Grant and G. Hailu,
  %``Brane couplings from bulk loops,''
  Phys.\ Lett.\ B {\bf 506} (2001) 207,
  [arXiv:hep-ph/0012379].
\bibitem{rad_cor_cheng}H. C. Cheng, K. T. Matchev and M. Schmaltz,  
%``Radiative corrections to Kaluza-Klein masses,''  
Phys.\ Rev.\ D {\bf 66} (2002) 036005,  
[arXiv:hep-ph/0204342]. 
\bibitem{Dvali}
  G. R. Dvali, G. Gabadadze, M. Kolanovic and F. Nitti,
  %``The Power of brane induced gravity,''
Phys. Rev. D {\bf 64} (2001) 084004, [arXiv:hep-ph/0102216].
\bibitem{carena}  
%\cite{Carena:2002me} 
%\bibitem{Carena:2002me} 
 M. S. Carena, T. M. P. Tait and C. E. M. Wagner, 
 %``Branes and orbifolds are opaque,'' 
 Acta Phys. Polon. B {\bf 33} (2002) 2355, [arXiv:hep-ph/0207056]. 
 %%CITATION = HEP-PH/0207056;%% 

\bibitem{delAguila}
  F. del Aguila, M. Perez Victoria and J. Santiago,
  %``Some consequences of brane kinetic terms for bulk fermions,''
JHEP {\bf 0302} (2003) 051, [arXiv:hep-th/0302023];  F. del Aguila, M. Perez Victoria and J. Santiago, [arXiv:hep-ph/0305119].
  %%CITATION = HEP-PH/0305119;%%
%\cite{delAguila:2003bh}
%  F.~del Aguila, M.~Perez-Victoria and J.~Santiago,
  %``Bulk fields with general brane kinetic terms,''.
%%CITATION = HEP-TH/0302023;%%

\bibitem{delAguila_STU}
  F. del Aguila, M. Perez-Victoria and J. Santiago,
  %``Physics of brane kinetic terms,''
  Acta Phys. Polon. B {\bf 34} (2003) 5511,
  [arXiv:hep-ph/0310353].

\bibitem{schwinn} 
%\cite{Schwinn:2004xa} 
%\bibitem{Schwinn:2004xa} 
 C. Schwinn, 
 %``Higgsless fermion masses and unitarity,'' 
 Phys. Rev. D {\bf 69} (2004) 116005, [arXiv:hep-ph/0402118]. 
 %%CITATION = HEP-PH/0402118;%% 


\bibitem{flacke}  
%\cite{Flacke:2008ne} 
%\bibitem{Flacke:2008ne} 
 T.~Flacke, A.~Menon and D.~J.~Phalen, 
 %``Non-minimal universal extra dimensions,'' 
 Phys.\ Rev.\ D {\bf 79} (2009) 056009, [arXiv:0811.1598 [hep-ph]]. 
 %%CITATION = ARXIV:0811.1598;%%%\cite{Schwinn:2004xa} 

\bibitem{ddrs1} 
%\cite{Datta:2012xy} \bibitem{Datta:2012xy} 
A. Datta, U. K. Dey, A. Shaw and A. Raychaudhuri, 
%``Universal Extra-Dimensional Models with Boundary localised Kinetic Terms: Probing at the LHC,''
 Phys. Rev. D {\bf 87} (2013) 076002, [arXiv:1205.4334 [hep-ph]].
\bibitem{flacke_STU} T. Flacke, K. Kong and S. C. Park, Phys. Lett. B {\bf 728} (2014) 262, [arXiv:1309.7077 [hep-ph]].
\bibitem{tommy}J. Bonnevier, H. Melbeus, A. Merle and T. Ohlsson, Phys. Rev. D {\bf 85} (2012) 043524, [arXiv:1104.1430 [hep-ph]].
\bibitem{ddrs2}A. Datta, U. K. Dey, A. Raychaudhuri and A. Shaw,
%``Boundary localised Terms in Universal Extra-Dimensional Models through a Dark Matter perspective,''
 Phys. Rev. D {\bf 88} (2013) 016011, [arXiv:1305.4507 [hep-ph]].
\bibitem{zbb}T. Jha and A. Datta, JHEP {\bf 1503} (2015) 012, [arXiv:1410.5098 [hep-ph]].
\bibitem{tirtha}U. K. Dey and T. S. Roy, Phys. Rev. D {\bf 88} (2013) 056016, [arXiv:1305.1016 [hep-ph]]. 
 
\bibitem{asesh_lhc1}
%\cite{Datta:2012at}
%\bibitem{Datta:2012at}
A. Datta, K. Nishiwaki and S. Niyogi,
  %``Non-minimal Universal Extra Dimensions: The strongly interacting sector at %the Large Hadron Collider,''
JHEP {\bf1211} (2012) 154, [arXiv:1206.3987 [hep-ph]],
%%CITATION = ARXIV:1206.3987;%%
%\bibitem{asesh_lhc2}
A. Datta, K. Nishiwaki and S. Niyogi,
  %``Non-minimal Universal Extra Dimensions: The strongly interacting sector at %the Large Hadron Collider,''
JHEP {\bf1401} (2014) 104, [arXiv:1310.6994 [hep-ph]].

\bibitem{lhc}A. Datta, A. Raychaudhuri and A. Shaw, Phys. Lett. B {\bf{730}} (2014) 42, [arXiv:1310.2021 [hep-ph]]; A. Shaw, Eur. Phys. J. C {\bf 75} (2015) 33, [arXiv:1405.3139 [hep-ph]].
%\bibitem{gamma_exp}
%\bibitem{gamma_sm}
\bibitem{cms}S.\;Chatrchyan et\;al, CMS Collaboration, Phys. Rev. Lett, {\bf 111} (2013) 101804, [arXiv:1307.5025[hep-ex]].
\bibitem{lhcb}R.\;Aaij et\;al, LHCb Collaboration, Phys. Rev. Lett, {\bf 111} (2013) 101805, [arXiv:1307.5024[hep-ex]].
\bibitem{sm} C. Bobeth, et al. Phys. Rev. Lett, {\bf 112} (2014) 101801, [arXiv:1311.0903[hep-ph]].
\bibitem{buras}A. J. Buras, M. Spranger and A. Weiler, Nucl. Phys. B {\bf 660} (2003) 225, [arXiv:hep-ph/0212143].
%; A. J. Buras, A. Poschenrieder, M. Spranger and A. Weiler, Nucl. Phys. B {\bf 678} (2004) 455, [arXiv:hep-ph/0306158].
%\bibitem{muong2}T. Appelquist B. A. Dobrescu Phys. Lett. B {\bf 516} (2001) 85, [arXiv:hep-ph/0106140].
\bibitem{gf} 
%\cite{Datta:2012xy} \bibitem{Datta:2012xy} 
A. Datta and A. Shaw, [arXiv:1408.0635 [hep-ph]].



%\bibitem {rho1} I. Gogoladze and C. Macesanu, Phys. Rev. D {\bf 74} (2006) 093012 [arXiv:hep-ph/0605207].
%\bibitem {rho2} M. Baak et al., Eur. Phys. J. C {\bf 72} (2012) 2003 [arXiv:1107.0975].
%\bibitem {pt_para} M.E. Peskin and T. Takeuchi, Phys. Rev. Lett. {\bf 65} (1990) 964;  M.E. Peskin and T. Takeuchi, Phys. Rev. D {\bf 46} (1992) 381.

\bibitem{ntr}Nature {\bf 522} (2015) 68, [arXiv:1411.4413[hep-ex]].
\bibitem {pdg}K. A. Olive et al. (Particle Data Group), Chin. Phys. C, {\bf 38} (2014) 090001. 
\bibitem {kksum} A. Datta and S. Raychaudhuri, Phys. Rev. D {\bf 87} (2013) 035018, [arXiv:1207.0476 [hep-ph]].
\bibitem{gg_converge}G. Bhattacharyya and P. Dey, Phys. Rev. D {\bf 70} (2004) 116012, [arXiv:hep-ph/0407314].

%\bibitem {hgs} A. Datta, A. Patra and S. Raychaudhuri, Phys. Rev. D {\bf 89} (2014) 093008, [arXiv:1311.0926 [hep-ph]].
\bibitem {mu2} P. Nath, M. Yamaguchi, Phys. Rev. D {\bf 60} (1999) 116006, [arXiv:hep-ph/9903298].
\bibitem {rho} T. Appelquist and H. U. Yee, Phys. Rev. D {\bf 67} (2003) 055002, [arXiv:hep-ph/0211023].

\bibitem{buras3}A. J. Buras, A. Poschenrieder, M. Spranger and A. Weiler, Nucl. Phys. B {\bf 678} (2004) 455, [arXiv:hep-ph/0306158].
\bibitem {fcnc} K. Agashe, N. G. Deshpande and G. H. Wu, Phys. Lett. B {\bf 514} (2001) 309, [arXiv:hep-
ph/0105084]; D. Chakraverty, K. Huitu and A. Kundu, Phys. Lett. B {\bf 558} (2003) 173,
[arXiv:hep-ph/0212047].
\bibitem {elctrowk} A. Strumia, Phys. Lett. B {\bf 466} (1999) 107, [arXiv:hep-ph/9906266]; T. G. Rizzo and J.
D. Wells, Phys. Rev. D {\bf 61} (2000) 016007, [arXiv:hep-ph/9906234]; C. D. Carone, Phys.
Rev. D {\bf 61} (2000) 015008, [arXiv:hep-ph/9907362].
\bibitem{bsg} U.~Haisch and A.~Weiler, Phys.\ Rev.\ D {\bf 76} (2007), 034014, [arXiv:hep-ph/0703064].
\bibitem {belayev} A. Belyaev, M. Brown, J. M. Moreno and C. Papineau, JHEP {\bf 1306} (2013) 080, [arXiv:1212.4858 [hep-ph]]. 

\bibitem {flacke_kong}T. Flacke, K. Kong and S.C. Park, JHEP {\bf 05} (2013) 111 [arXiv:1303.0872].
\bibitem {flacke_Pasold} T. Flacke and C. Pasold, Phys. Rev. D {\bf 85} (2012) 126007 [arXiv:1111.7250].
%\bibitem{gfit}M. Baak, R. Kogler (for the Gfitter group), [arXiv:1306.0571[hep-ph]]. 
\bibitem{gfit}M. Baak et al, (for the Gfitter group), Eur. Phys. J. C {\bf 74} (2014) 3046, [arXiv:1407.3792 [hep-ph]].

\end{thebibliography}
\end{document}